# Load-Dependent Sliding behavior of $WSe_{2-x}$ solid lubricant coating

Yue Wang[a,*], Himanshu Rai[a], Tomas Polcar[a,b,*]

a) Advanced Materials Group, Faculty of Electrical Engineering, Czech Technical University in Prague, Technicka 4, Prague 6, 16000, Czech Republic
b) School of Engineering, University of Southampton, Highfield, SO17 1BJ, Southampton, UK

**Abstract**

Transition-metal dichalcogenides (TMDs) are commonly used as solid lubricants in various environments. Molybdenum disulfide is the most studied and applied TMD solid lubricant, but other members may have similar or even better sliding properties. Tungsten diselenide is one of the materials that has rarely been investigated in terms of tribological properties. This paper provides a comprehensive tribological characterization of substoichiometric tungsten diselenide and molybdenum disulfide coatings deposited by magnetron sputtering. We focused on tribological properties at a macroscopic scale, particularly friction and wear dependence on applied load; however, a nanoscale frictional assessment of worn surfaces was performed as well to identify the major wear mechanisms. Substoichiometric tungsten diselenide outperformed traditional molybdenum disulfide, exhibiting much lower friction in humid air, suggesting lower coating sensitivity to the humid atmosphere. Moreover, a combination of nanotribological experiments in the wear tracks with sliding under different environmental conditions suggests that the key factor causing frictional load-dependence (deviation from Amonton's law) is frictional heating of the surface.

## 1. Introduction

Friction and wear cause substantial energy loss and material degradation and ultimately contribute to the mechanical failure of various engineering components. Solid lubricants possess exceptional physical and chemical properties, allowing them to reduce friction and wear in harsh conditions such as wide temperature ranges or various sliding speeds, making them potentially more efficient than liquid lubricants[1–3]. However, the challenge remains to identify optimum solid lubricants that can consistently and effectively minimize friction and wear in various sliding environments. Transition-metal dichalcogenides (TMDs) stand out because of their layered structure and weak van der Waals bonding between the layers. Their structure facilitates interlayer shear, resulting in exceptionally low frictional properties of TMDs in vacuum [4]. However, when applied in industrial settings, TMD-based solid lubricant coatings often face environmental degradation. Water and oxygen molecules in the atmosphere play a detrimental role; water molecules are absorbed between TMDs layers, reducing the lubrication function at room temperature, and atmospheric oxygen leads to rapid oxidation [5,6]. Molybdenum disulfide, the most extensively studied TMD material, oxidizes at relatively low temperatures (approximately 300 °C) and exhibits a much higher coefficient of friction in ambient environments compared to vacuum [7,8]. To overcome the limitations of $MoS_2$, alternative TMD materials with a similar structure have been studied, such as $WS_2$ and $MoSe_2$. Replacement of molybdenum with tungsten enhances the oxidation-resistance of TMDs; it has been reported that tungsten disulfide can maintain its low coefficient of friction up to 500 ℃ [9]. Meanwhile, molybdenum diselenide is a better choice for a humid environment, providing a lower coefficient of friction than that of $MoS_2$; the reduction of friction can reach 50% [10] .

Considering properties of $WS_2$ and $MoSe_2$, tungsten diselenide, another member of the TMD family with a lamellar structure, is a promising choice with the potential to resist environmental attacks. The nanofriction characterization on $WSe_2$ monolayers in ambient air showed the lowest friction when compared to other tested TMDs flakes [11].

Some studies explored tungsten diselenide applied as a solid lubricant coating. Dominguez-Meister [12] and Evaristo [13] developed tungsten diselenide-based coating with a coefficient of friction as low as 0.08, which was lower than other TMD materials like $MoS_2$ and $WS_2$. Yet, the properties of $WSe_2$ as a solid lubricant are still almost unknown, particularly when compared to the most studied solid lubricant, $MoS_2$. In this work, a comprehensive tribological characterization consisting of macroscale and nanoscale tribological tests was conducted on tungsten diselenide coatings deposited by magnetron sputtering and compared to $MoS_2$ reference films to investigate their tribological behavior. Macroscale tribological tests were conducted under various normal loads, while load-dependence nanotribological tests were performed on the wear tracks. Our results indicate that the $WSe_{2-x}$ coating exhibits superior tribological performance in ambient environments compared to $MoS_2$ and displays load-adaptive sliding behavior. It is worth noting that the majority of sputtered TMDs are deficient in chalcogenide due to resputtering [14,15], but they are traditionally referred to as disulfides or diselenides.

## 2. Experiment Section

### 2.1 Sample preparation

$WSe_{2-x}$ and $MoS_2$ coatings were deposited by an AJA magnetron sputtering system with a 150 W DC power supply using pure $WSe_2$ and $MoS_2$ targets. Polished steel samples (WNr. 1.2379 steel), with a hardness of about 9 GPa, were selected as the substrates for coatings. Before the deposition, the substrate was sputter-cleaned under 50 W RF power for 30 minutes. A pure Cr layer (~150 nm) was deposited between the coating and substrate to improve adhesion. The deposition parameters are shown in Table 1. The coating thickness was measured by Zygo NewView 8000 profilometer.

Table 1 Deposition parameters for coatings

| **Sample Name** | **Power (W)** | **Base Vacuum (Pa)** | **Working Pressure (Pa)** | **Deposition Time (s)** | **Thickness (nm)** |
|---|---|---|---|---|---|
| $MoS_2$ | 150 | $5\times10^{-3}$ | 0.67 | 5400 | 1320 |
| $WSe_{2-x}$ | 150 | $5\times10^{-3}$ | 0.67 | 7200 | 1030 |

**2.2 Structure Characterization**

The coatings' morphology was characterized by SEM (MIRA3 XMU). Energy dispersive spectroscopy (EDS) determined the chemical composition. X-ray diffraction (Rigaku 3; Cu Kα) revealed the structure, whereas X-ray photoelectron spectroscopy was employed to identify chemical states; the samples were ion-etched for 5 min to remove surface contamination. The hardness and Young's modulus were measured by nanoindentation. After tribological testing, the wear tracks and ball scars were characterized by Raman (Horiba Xplora) with 532 nm laser excitation, and the data were compared with spectra taken from virgin samples. The morphologies of the wear tracks were characterized by a 3D White light optical profilometer (Zygo NV7200).

**2.3 Macroscale Tribological Testing**

Tribological tests were carried out using a Bruker UMT-2 tribometer in reciprocating mode at 1 Hz, with a stroke length of 3 mm. The test duration was 1000 s. The normal load was set to 1, 3, 5, 7, and 10 N, 10 mm AISI440C ball was used as a counterpart. The corresponding Hertz contact pressure is shown in Fig. S1. The steady-state coefficient of friction was obtained by taking an average after 300 laps of testing to avoid the running-in effect. The wear rate was calculated according to the following equation [16]:

$$K = \frac{V}{FL} \qquad \text{Equation (1)}$$

Where V is the worn volume (mm$^3$), F is the normal load (N), and L is the sliding distance (m). All macroscale experiments were carried out in an ambient environment: 25 °C and relative air humidity of 40%.

## 2.4 Nanotribological Characterization

Nanotribological characterization of coatings and wear tracks was performed by scanning probe microscope (Bruker, Dimension ICON-SPM) in contact mode in an ambient environment. Surface topographies and friction maps were carried out on a 2 μm ×2 μm area. A DLC-coated AFM probe (ContDLC, BudgetSensors, Bulgaria) , and normal loads were set to increment from 20 nN to 70 nN following the procedure in Ref. [17,18]. The lateral and normal spring constants of the cantilever were calibrated following the methods proposed by Sader [19] and Green et al. [20]. Calibration of lateral and normal forces was performed using the beam geometry method [21,22]. All AFM measurements were conducted in contact mode under consistent ambient conditions.

# 3. Experimental results

## 3.1 Coating Characterization

The chemical composition of coatings was characterized by EDX, as shown in Table 2. The corresponding morphologies of each coating are shown in Fig. S2. Relatively high oxygen content in the as-deposited coatings is typical for such deposition processes and originates from residual atmosphere and from targets, which are porous and prone to surface oxidation. The $WSe_{2-x}$ coating exhibited a significant deviation from stoichiometry, with a Se/W atomic ratio of 0.92, consistent with previous studies[23]. The loss of selenium in the deposition of $WSe_{2-x}$ coating is likely the result of the resputtering effect caused by heavy tungsten atoms [24]. $MoS_2$ exhibits an S/Mo ratio of 1.91, which is close to stoichiometric. The $WSe_{2-x}$ coating showed higher hardness (3.5 GPa) and Young's modulus (87.7 GPa) compared to the $MoS_2$ coating (0.5 and 25.4 GPa, respectively).

Table 2 Chemical composition and mechanical properties of as-deposited coatings

| | Mo or W | S or Se | O | S/Mo or Se/W | Hardness (GPa) | Young's modulus (GPa) |
|---|---|---|---|---|---|---|
| $MoS_2$ | 30.1 | 57.3 | 12.6 | 1.91 | 0.5 | 25.4 |
| $WSe_{2-x}$ | 47.7 | 44.1 | 8.2 | 0.92 | 3.5 | 87.7 |

X-ray diffraction spectra are shown in Fig. 1 (a). In the case of $MoS_2$, the diffraction peaks of (002) and (100) at 14.1° and 33.5° belong to 2H-$MoS_2$[24]. A small peak at 59.2° refers to (110) in 2H-$MoS_2$[25]. Long asymmetric tail of (100) peak towards higher 2θ suggests turbostrating stacking of 10 L planes, a typical feature of sputtered $MoS_2$ coatings [26]. $WSe_{2-x}$ pattern shows a peak located at 13.8°, which fits well (002) reflections of 2H-$WSe_2$ [27] A large, broad peak located at 35-45° shows likely bcc tungsten (110) in the form of nanograins; moreover, 10L reflections of $WSe_2$ (L=1, 2, …) may contribute to this peak as well, possibly with a turbstrating effect discussed above. The Raman spectrum shown in Fig. 1 (b) exhibits two significant regions in the spectrum of $WSe_{2-x}$ coating. Peak at 245 $cm^{-1}$ is very close to positions of $E^1_{2g}$ and $A_{1g}$ modes in 2H-$WSe_2$ [28] (note that the separation of these modes is almost negligible in $WSe_2$, about 8 $cm^{-1}$), suggesting that the 2H phase is predominant in the coating. A smaller peak is positioned close to 170 $cm^{-1}$, which can be identified as $E_{1g}$ in 2H-$WSe_2$ [29]. Broad peaks at 807 $cm^{-1}$ belong to $WO_3$, suggesting a small amount of oxidation in the coating, as already indicated by EDX measurement [30]. The Raman spectrum of $MoS_2$ shows peaks at 169 $cm^{-1}$, 371 $cm^{-1}$, and 405 $cm^{-1}$, fitting well with the typical Raman spectrum of 2H-$MoS_2$ [31]. Small peaks in the region of 750-930 $cm^{-1}$ indicate the presence of oxide phases.

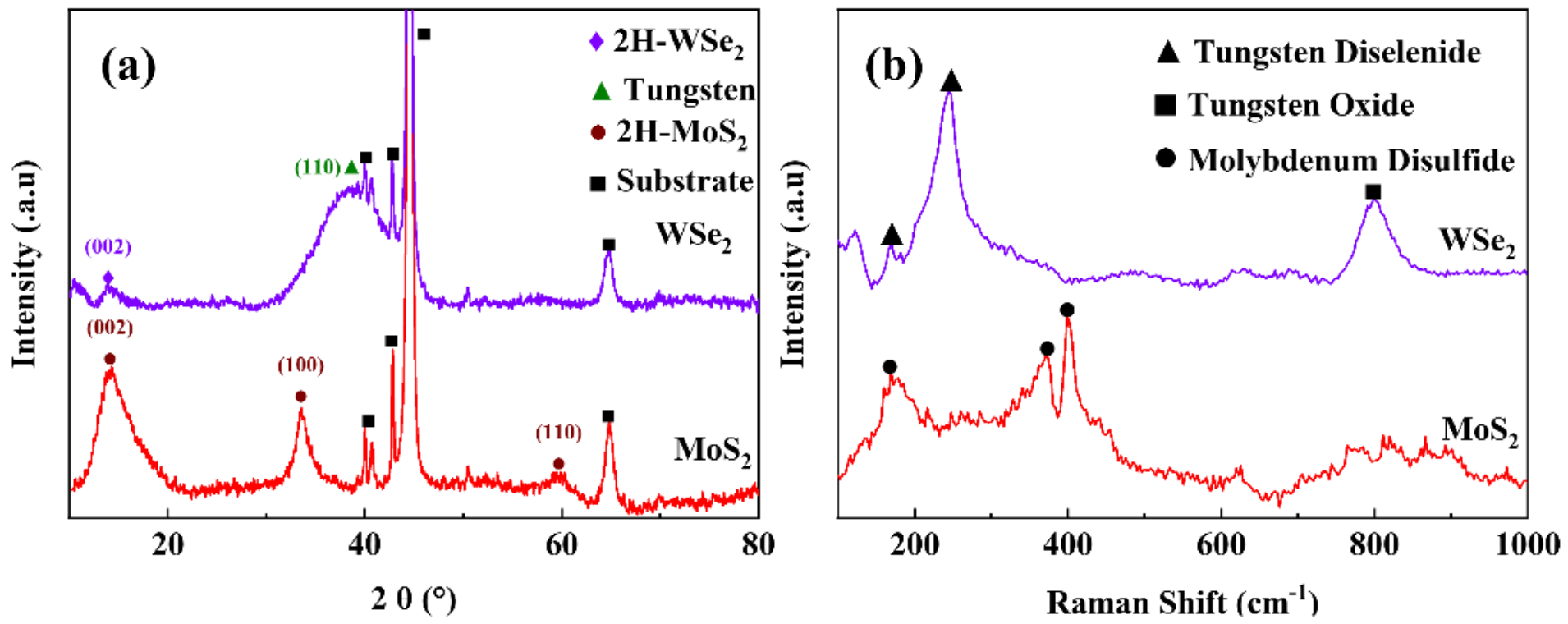

Fig. 1 Structural characteristics of as-deposition coatings: (a) XRD, (b) Raman spectroscopy

The X-ray photoelectron spectroscopy (XPS) was conducted on each coating to assess the surface chemistry state further. As illustrated in Fig. 2(a), the W 4f spectrum in $WSe_{2-x}$ coatings contains $W^{4+}$, associated with $WSe_2$, $W^{6+}$, attributed to tungsten oxides, and $W^0$ corresponding with metallic state tungsten. The dual peaks at 32.2 eV and 34.1 eV correspond to the $4f_{5/2}$ and $4f_{7/2}$ levels of $WSe_2$, respectively[32]. Due to oxidation, sharp peaks are observed at 35.6 eV and 37.7 eV, which are linked to tungsten oxides ($WO_x$) [33,34]. Additionally, a small peak at 41.8 eV corresponds to W 5p, also associated with $WO_x$ [35]. The peaks representing metallic tungsten can also be indexed at 31.4 eV and 33.4 eV [36]. Aside from the characteristic twin peaks at 54.7 eV and 55.5 eV for $WSe_2$ [37], another pair of peaks at 53.8 eV and 54.4 eV can is present. These extra peaks are often attributed to $Se^{2-}$, shifted likely due to the presence of oxygen combined with low selenium content in $WSe_{2-x}$ coating [38]. O1s spectrum shows a broad peak at 531.3 eV, which fits $O^{2-}$ in Se-O binding, while a W-O binding peak can also be recognized at 530.8 eV [39,40]. In the case of the as-deposited $MoS_2$ coating, $Mo^{4+}$ is present in $MoS_2$, and $Mo^{6+}$ is linked to molybdenum oxides. Two prominent peaks at 229.2 eV and 232.4 eV correspond to the Mo $3d_{5/2}$ and Mo $3d_{3/2}$ levels of $MoS_2$, respectively [41]. A small peak at 235.3 eV, corresponding to the Mo-O bond, indicates surface oxidation of the $MoS_2$ coating [42]. Two types of oxides can be detected in the $MoS_2$ coating from O1s spectrum (Fig. 2 (f)): $MoO_2$ with $O^{2-}$ peak at 531.5 eV, and $MoO_3$ with $O^{2-}$ peak positioned at 530.5 eV [43,44]. Note that metallic molybdenum was not observed, which corresponds well with a high stoichiometry of this coating.

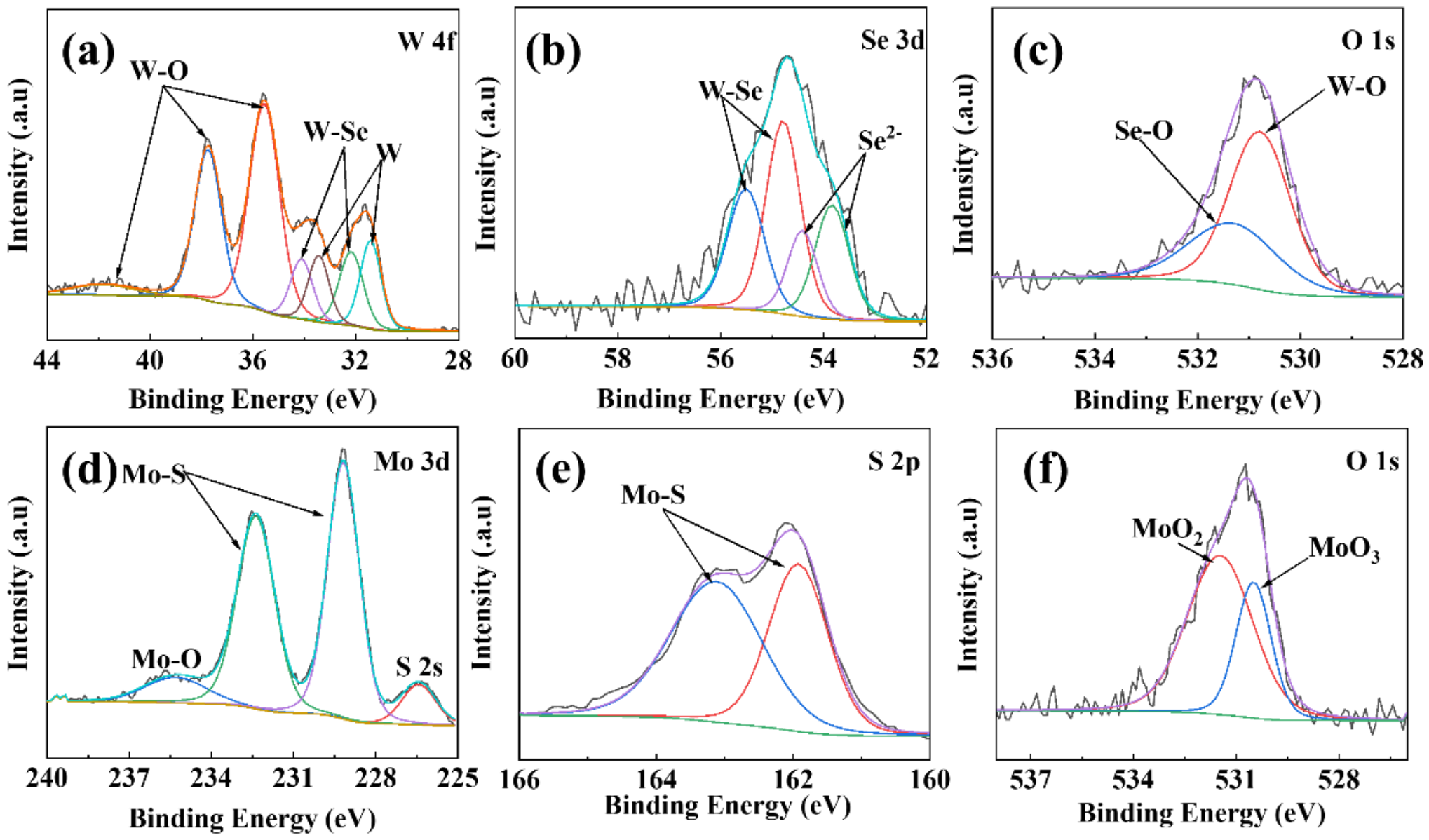

Fig. 2 X-ray photoelectron spectrum of $WSe_{2-x}$: (a) W 4f, (b) Se 3d, (c) O 1s; and of $MoS_2$: (d) Mo 3d, (e) S 2p, (f) O 1s

Fig. 3 shows surface topography and friction force maps of coatings. $WSe_{2-x}$ coating exhibits a compact cluster-like morphology with a low surface roughness Ra of around 4 nm. $MoS_2$ coating surface, in contrast, includes many holes (Fig. 3(c)), contributing to rougher morphology with a higher surface roughness (Ra~28.8 nm).

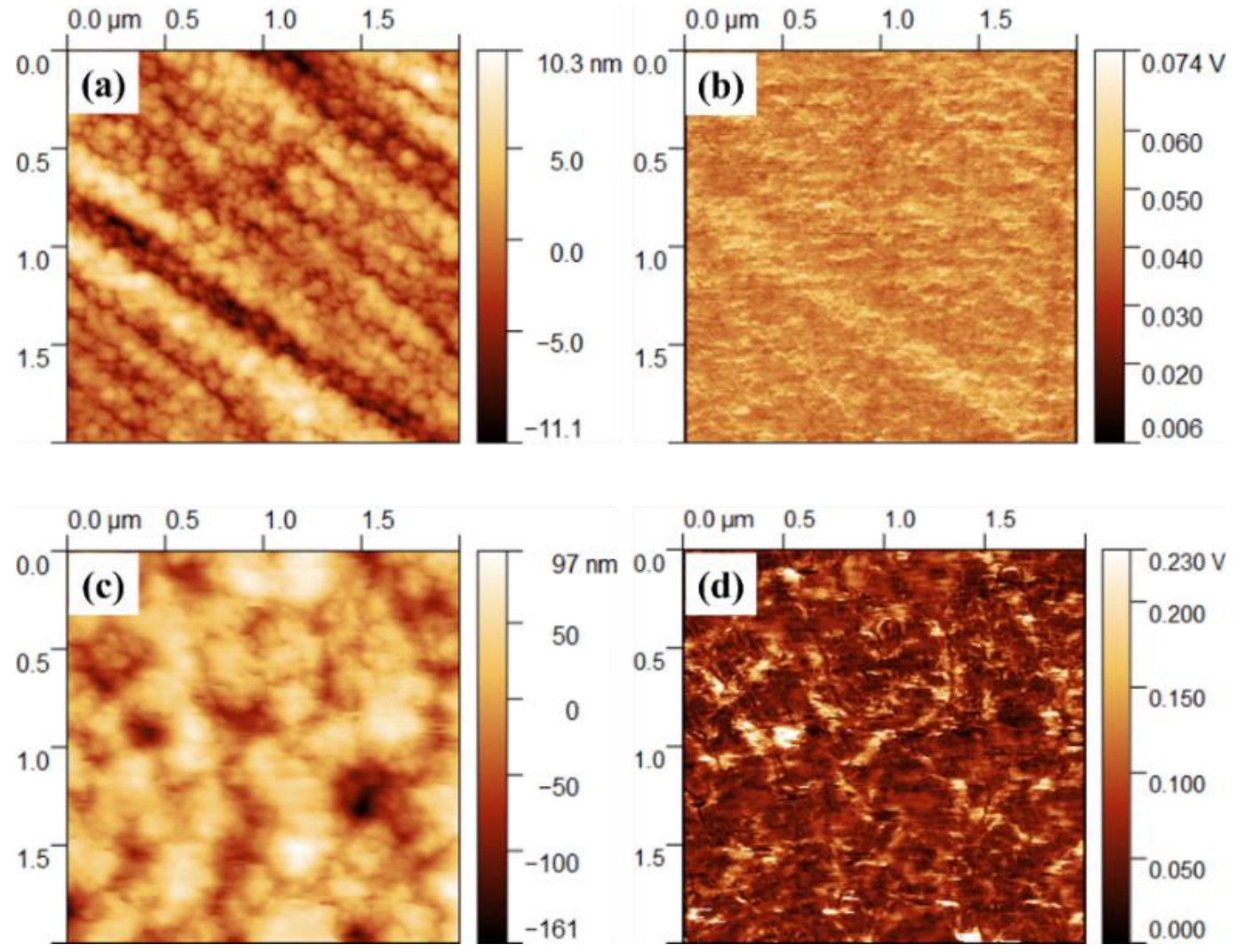

Fig. 3 Surface topographies of $WSe_{2-x}$ (a) and $MoS_2$ (c) and corresponding friction

force maps: $WSe_{2-x}$ (b) and $MoS_2$ (d)

### 3.2 Macroscale tribological testing: Load dependence

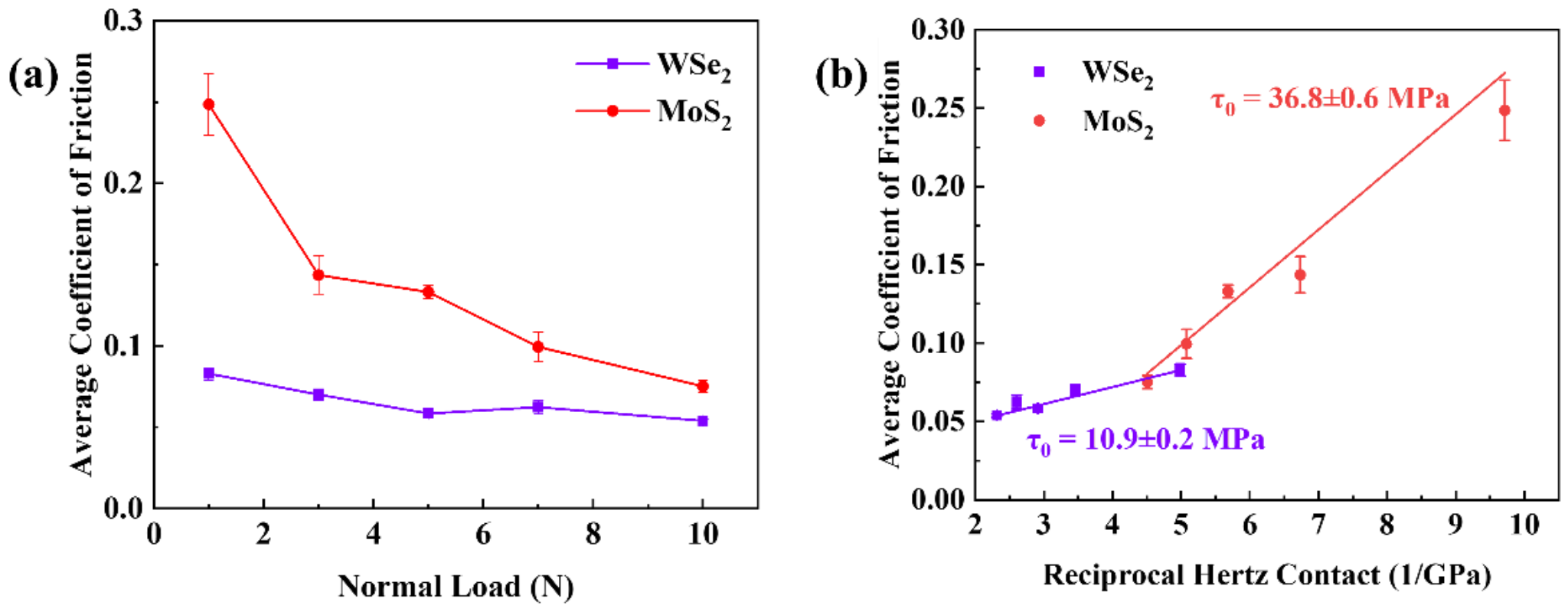

Fig. 4 (a) The average coefficient of friction of $WSe_2$ and $MoS_2$ coatings and (b) linear regression fit of steady-state coefficient of friction of $WSe_2$ and $MoS_2$.

The frictional properties of the as-deposited coatings were evaluated using a UMT-2 tribometer under ambient conditions. The friction curves are shown in Fig. S3. As shown in Fig. 4, $WSe_{2-x}$ coatings demonstrated a very low coefficient of friction in the ambient environment, decreasing with increasing normal load from 0.083 to 0.054. $MoS_2$ follows the same trend, but with much higher values in a range of 0.248 – 0.075. The average coefficient of friction of TMD solid lubricants is often fitted using Hertz contact [45]:

$$\mu = \tau_0 \pi (\frac{3R}{4E})^{2/3} L^{-2/3} + \alpha, \qquad \text{Equation (2)}$$

Where $\mu$ is the coefficient of friction, $\tau_0$ is the shear strength at the tribological interface, R is the radius of the ball, E is Young's modulus, L is the normal load, and $\alpha$ is the coefficient representing the coefficient of friction at zero loads (component related mostly to surface adhesion). As shown in Fig 4 (b), the coefficient of friction of both coatings fits well the Equation 1 with $\tau_0$ of $WSe_{2-x}$ and $MoS_2$ coatings calculated as 10.9 ±0.2 and 36.8±0.6 MPa, respectively. The value for $MoS_2$ corresponds well with the literature [46].

3D topography of the wear tracks are presented in Fig. S4, with corresponding optical images provided in Fig. S5. Pronounced grooves were observed on all wear tracks of the $WSe_{2-x}$ coatings, particularly at 7 N and 10 N. By contrast, the loose structure of the $MoS_2$ coating was compressed during testing, resulting in a flat and smooth worn surface.

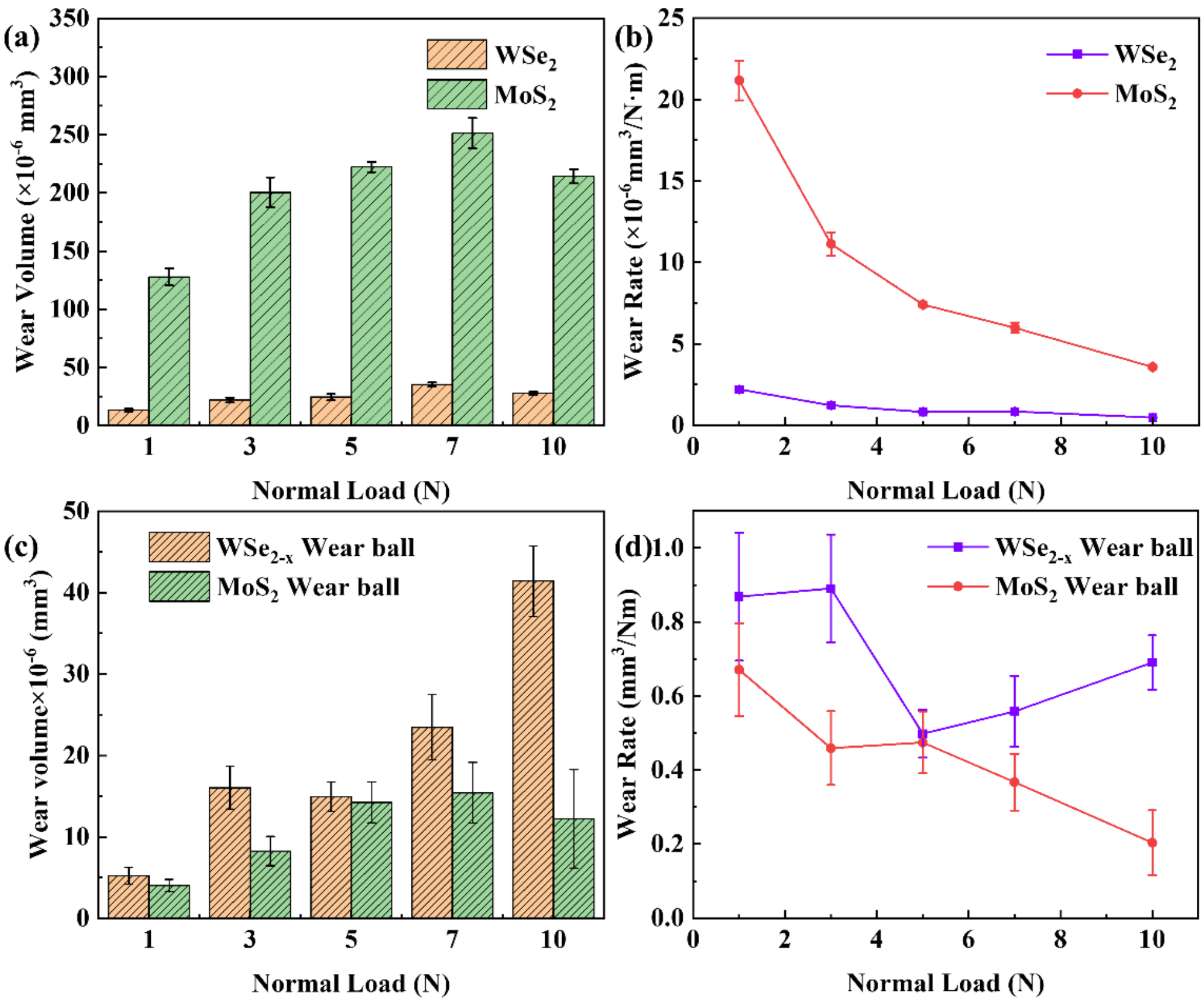

Fig. 5 (a) The worn volumes and (b) corresponding wear rate of $WSe_{2-x}$ and $MoS_2$ coatings; (c) the worn volumes and (d) corresponding wear rate of $WSe_{2-x}$ and $MoS_2$ wear balls.

The worn volumes and corresponding wear rates are shown in Fig. 5. $WSe_{2-x}$ coating wear volume increases with the normal load up to 7 N; then it slightly decreases, a behavior observed in other TMD-based coatings as well [47]. $MoS_2$ coating showed a similar trend, an increase in wear volume up to a maximum at a normal load of 7 N. As a consequence, the wear rate of both coatings decreases with the applied load, but the wear rate of $WSe_{2-x}$ is significantly lower in the entire load range. The wear on the counterparts is very low for both coatings, with wear rates below $1\times10^{-6}$ $mm^3/Nm$ as shown in Fig. 5 (d). The balls sliding against $WSe_{2-x}$ exhibit higher wear rates than those sliding against $MoS_2$; nevertheless, the ball wear shows different trends for the two solid lubricant coatings. Unlike continuously decreasing $MoS_2$ ball wear with load, $WSe_{2-x}$ ball wear shows more or less random evolution independent of normal load.

Fig. 6 shows Raman spectra taken from the wear track. $MoS_2$ underwent significant recrystallization during sliding. Distinct peaks at around 400 $cm^{-1}$, corresponding to the 2H-$MoS_2$ phase, are present in all post-test spectra. Moreover, the peaks ($E_{2g}$ and $A_{1g}$) slightly shifted from an initial (unworn surface) position when acquired from the wear track, but the peak position remained almost independent of the applied load. However, it is worth noting that the peaks of the as-deposited surface are not well defined, and identification of their position is difficult. $WSe_2$ spectra from the wear tracks were very similar to that of the as-deposited coating, and the shift of $E_{12g}$ and $A_{1g}$ peaks shown in Fig. S6 is quite random, although always moving into higher peak positions. Higher values of $E_{1g}$ peak indicate a different structure of the worn surface [44,45], but our Raman spectra are not conclusive to state that a $WSe_2$ tribolayer with a higher degree of crystallinity has been formed.

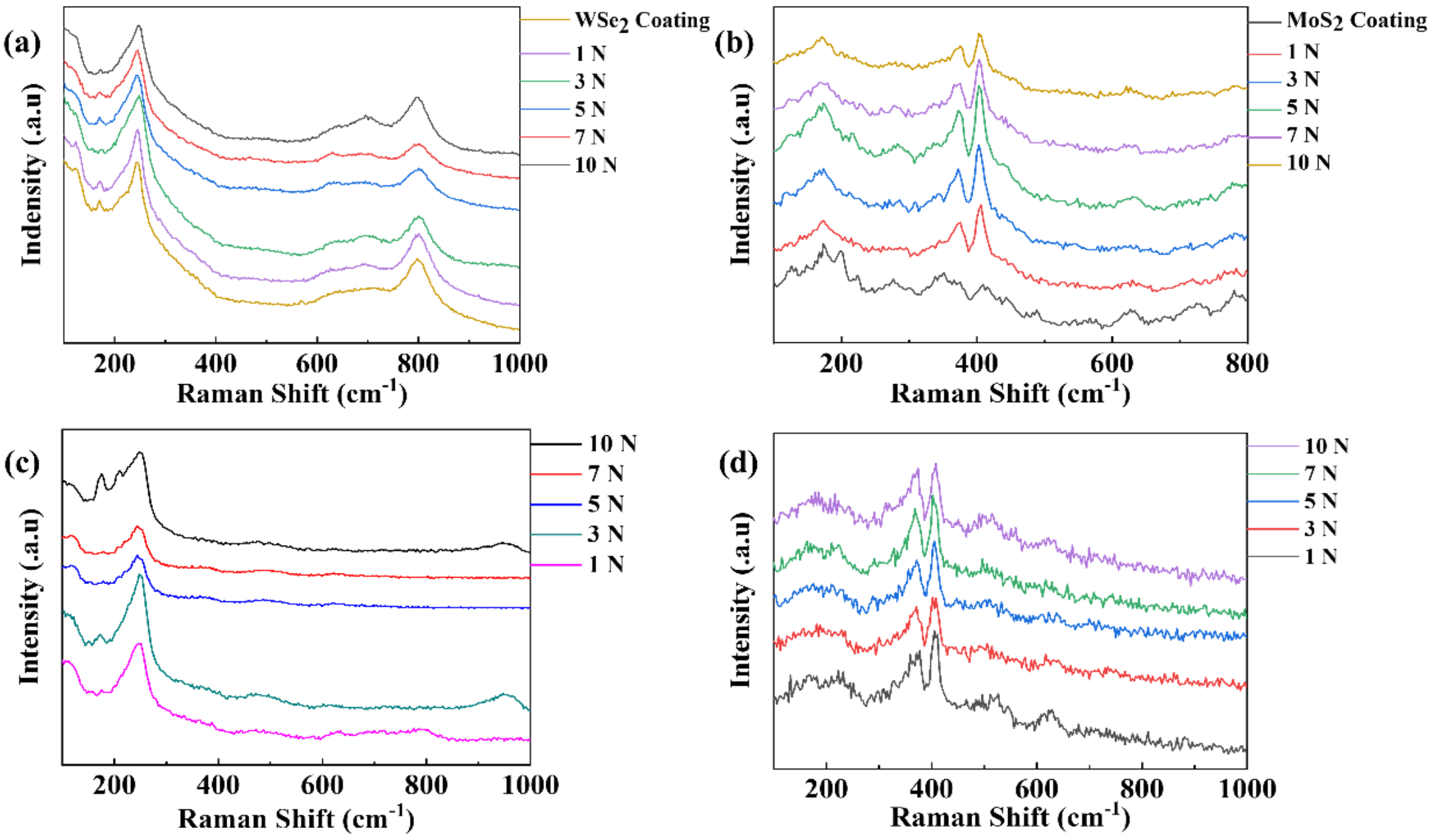

Fig. 6 Raman spectra on the wear tracks: (a) $WSe_{2-x}$ coating, (b) $MoS_2$ coating; and on the ball scars: (c) $WSe_{2-x}$, (d) $MoS_2$

As expected, a friction-induced material transfer from the wearing coating to the counterparts (balls) occurred for both solid lubricants (Fig. S7 and Fig. S8). Raman spectra taken from the center of transferred materials adhering to ball scars are shown in Fig. 6. Both $WSe_{2-x}$ and $MoS_2$ show significant Raman peaks for 2H-$WSe_2$ and 2H-$MoS_2$ for all loads. Note that there is almost no oxidation, as demonstrated by the absence of peaks at ~820 $cm^{-1}$ for Mo-$O_x$ and ~810 $cm^{-1}$ for W-$O_x$ [30,48]. Moreover, there is no evidence of Raman peak in positions close to iron oxides or more complex iron-based compounds (Table 3). These results indicate that the ball wear is likely caused during running in, and then the wear scar is covered (and protected) by material transferred from the solid lubricant coatings. The position of peaks shown in Fig. S9 is very similar to that of the wear track. Combining the morphology and Raman spectra, the $WSe_{2-x}$ shows a lower amount of coating material transferred to the ball when compared to $MoS_2$, which can contribute to slightly higher ball wear during sliding with $WSe_{2-x}$ coating. Nevertheless, we can conclude here that the ball wear is negligible in both cases, limited to the initial running-in stage, and does not significantly influence the tribological properties of the coatings.

Table 3 Raman peak positions of potential iron-based oxide tribochemical products.

| Phases | Raman peak positions ($cm^{-1}$) |
|---|---|
| $Fe_2O_3$ | 229, 294, 410 $cm^{-1}$ [49] |
| $Fe_3O_4$ | 532, 667 $cm^{-1}$ [49] |
| β-$FeMoO_4$ | 943$cm^{-1}$[50] |
| $FeWO_4$ | 878 $cm^{-1}$ [51,52] |

Fig. 7 provides a more precise representation of the wear track surface structure evolution through the ratio of peak intensities. The tribo-active structure change is profound on $MoS_2$ coatings as the ratio of $I(A_{1g})/I(E_{2g})$ is much higher on transfer layers and wear tracks when compared to the as-deposited coating. Thus, it is clear evidence of the formation of a highly crystalline tribolayer. As the $I(A_{1g})/I(E_{2g})$ ratio does not change with increasing normal load, it is evident that the transformation of the wearing surface is completed already at the lowest load. The peak intensity ratio $I(E_{1g})/I(E^1_{2g}+A_{1g})$ is almost identical to as-deposited coating surface in the case of $WSe_{2-x}$ coatings; it increases only in the case of the highest load.

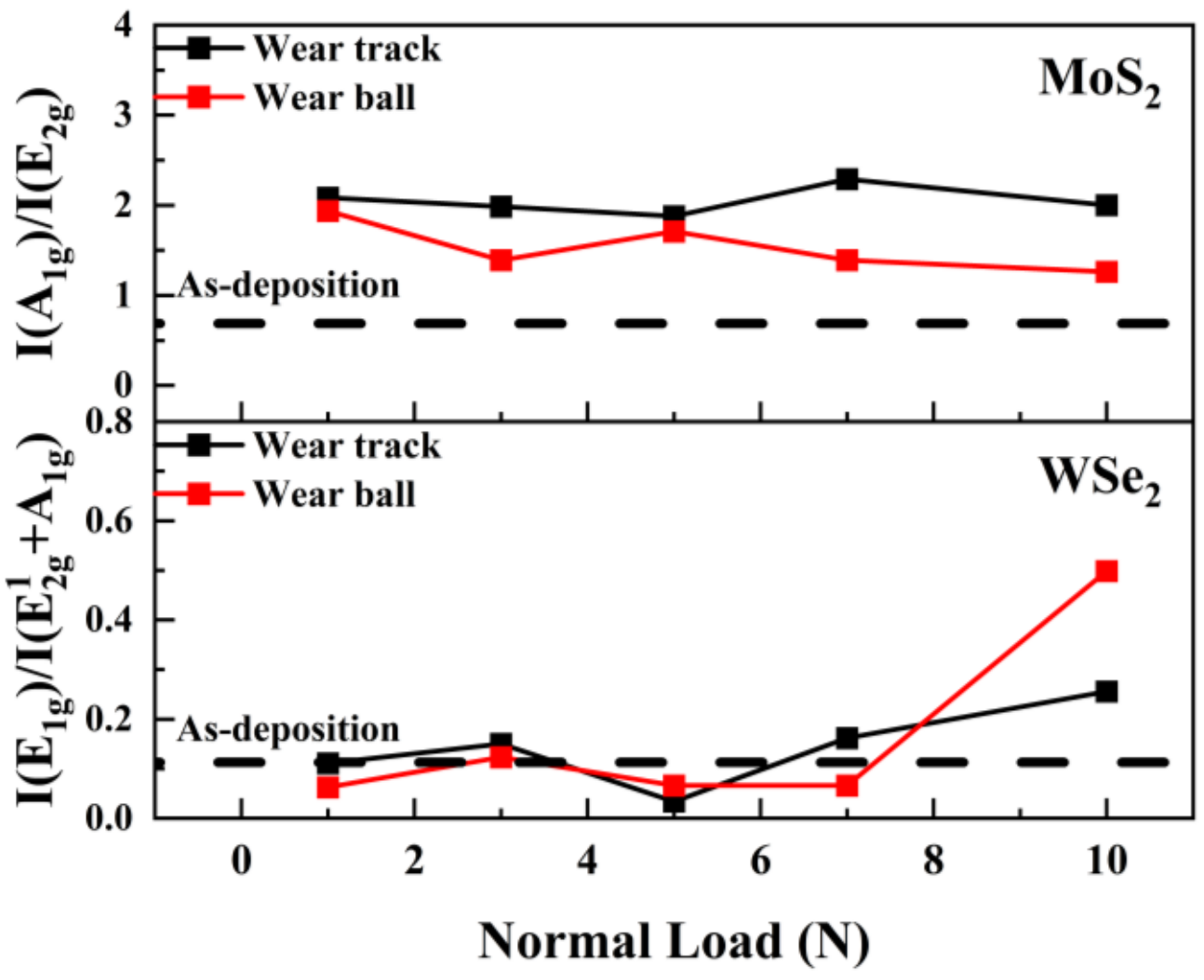

Fig. 7 The intensity ratios of the $A_{1g}$ and $E_{2g}$ peaks for $MoS_2$, and the $E_{1g}$ and $(E^1_{2g} + A_{1g})$ peaks for $WSe_2$

### 3.3 Nanotribological testing: wear track analysis

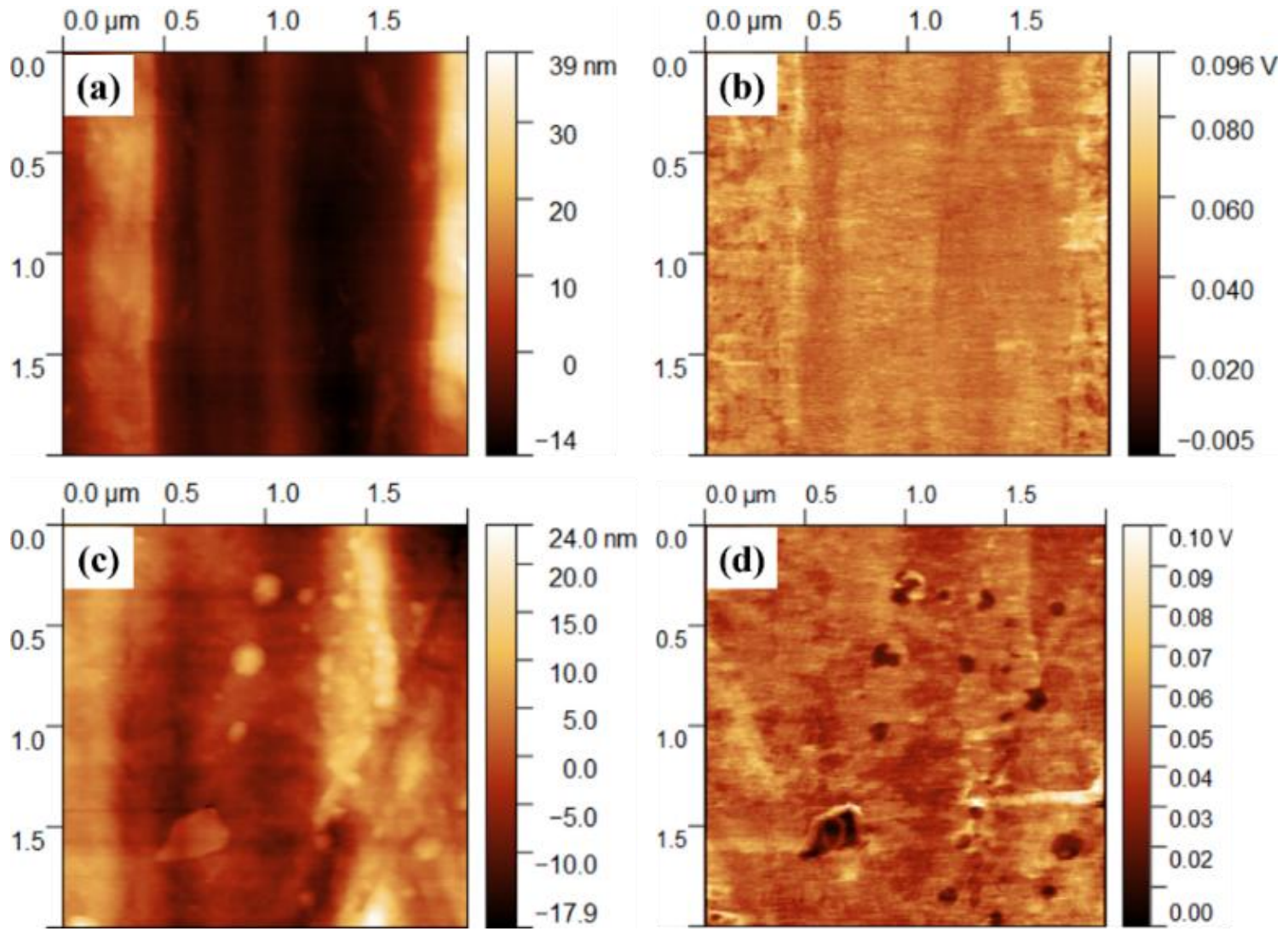

Fig. 8 Topographies and friction maps of wear tracks on $WSe_{2-x}$ sample: (a) and (b) 1 N, (c) and (d) 10 N

Nanotribological test was deployed on both TMD-based coatings to evaluate the tribological mechanism of $WSe_{2-x}$ coating. Nanotribological testing and surface topography were collected by SPM on 1 N and 10 N wear tracks. A 2 μm × 2 μm region was selected at the center of the wear track to avoid any macroscale defects. Fig. 8 shows topographies of wear tracks for 1 N and 10 N testing on the $WSe_{2-x}$ sample. Roughness is slightly higher than for unworn coating: Ra~8.2 nm for 1 N wear track and Ra~5.4 nm for 10 N wear track. The friction map in Fig. 8 (b) is uniform and independent of topography under low contact stress. Meanwhile, the topography of the 10 N wear track showed several small (50-300 nm) particles adhered on the surface, which exhibited significantly lower friction than the rest of the wear track (Fig. 8 (d)). These particles are almost identical in size and friction to those observed by Rapuc et al [53].

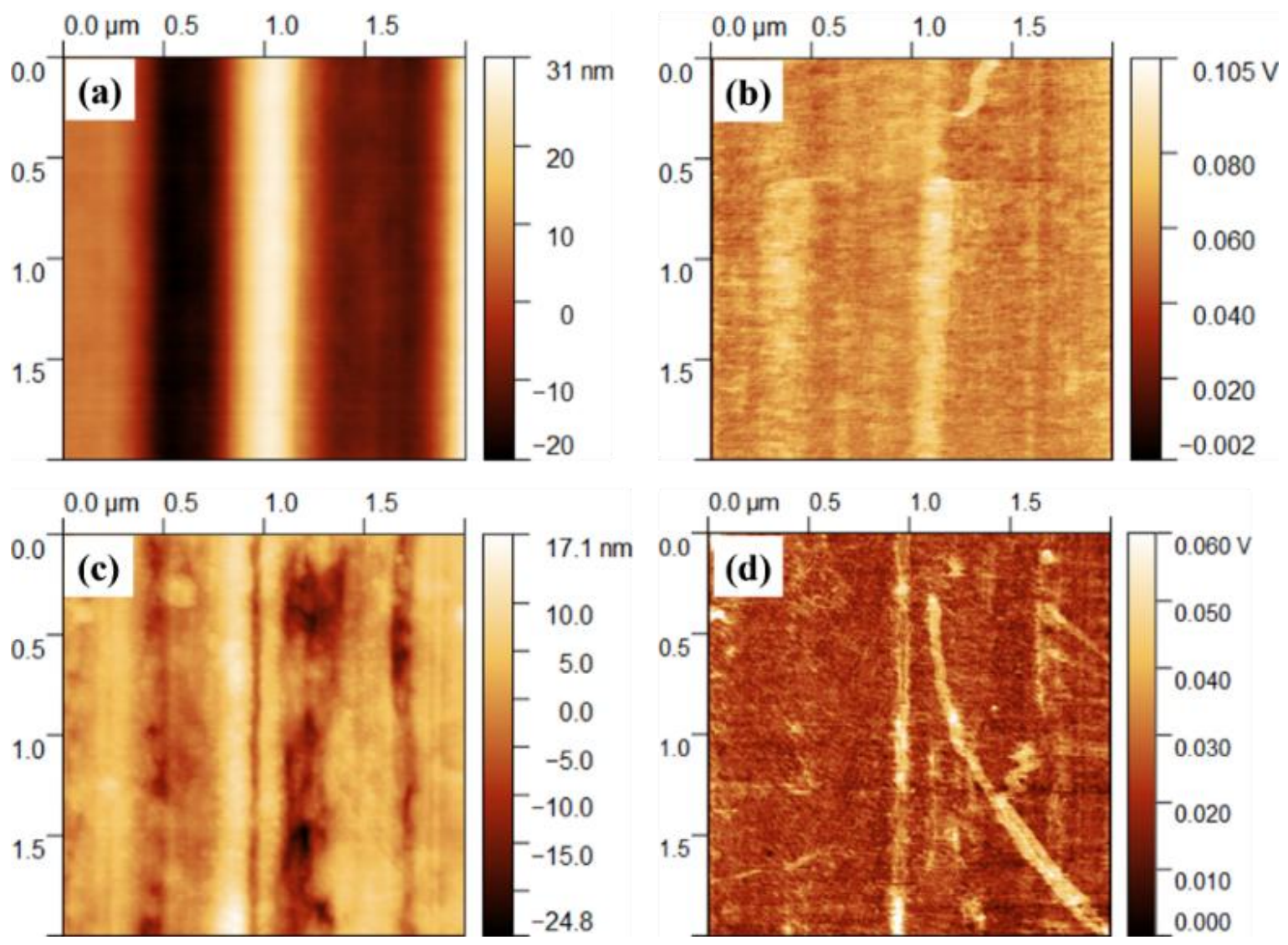

Fig. 9 Topographies and friction maps of wear tracks on $MoS_2$ sample: (a) and (b) 1 N, (c) and (d) 10 N

Compared with the $WSe_{2-x}$ sample, sputtered $MoS_2$ shows uniform friction maps, and no particles were observed at any of the worn surfaces, see Fig. 9. At the highest load (10N), the surface was noticeably polished with Ra~4.8 nm.

Finally, a load-dependent friction testing was also carried out on the as-deposited coating surface and inside the selected wear tracks (region of 2 μm×2 μm). The coefficient of friction was significantly lower inside all wear tracks, as shown in Fig. 10. Moreover, the coefficient of friction on the wear tracks produced at 10 N load is lower than that of wear tracks produced with 1N load. It is worth noting that $MoS_2$ coatings exhibit a much higher drop in friction when measured outside and inside the wear tracks, suggesting strong structural (crystalline tribolayer formation) and topographical (polishing) transformation of the worn surface.

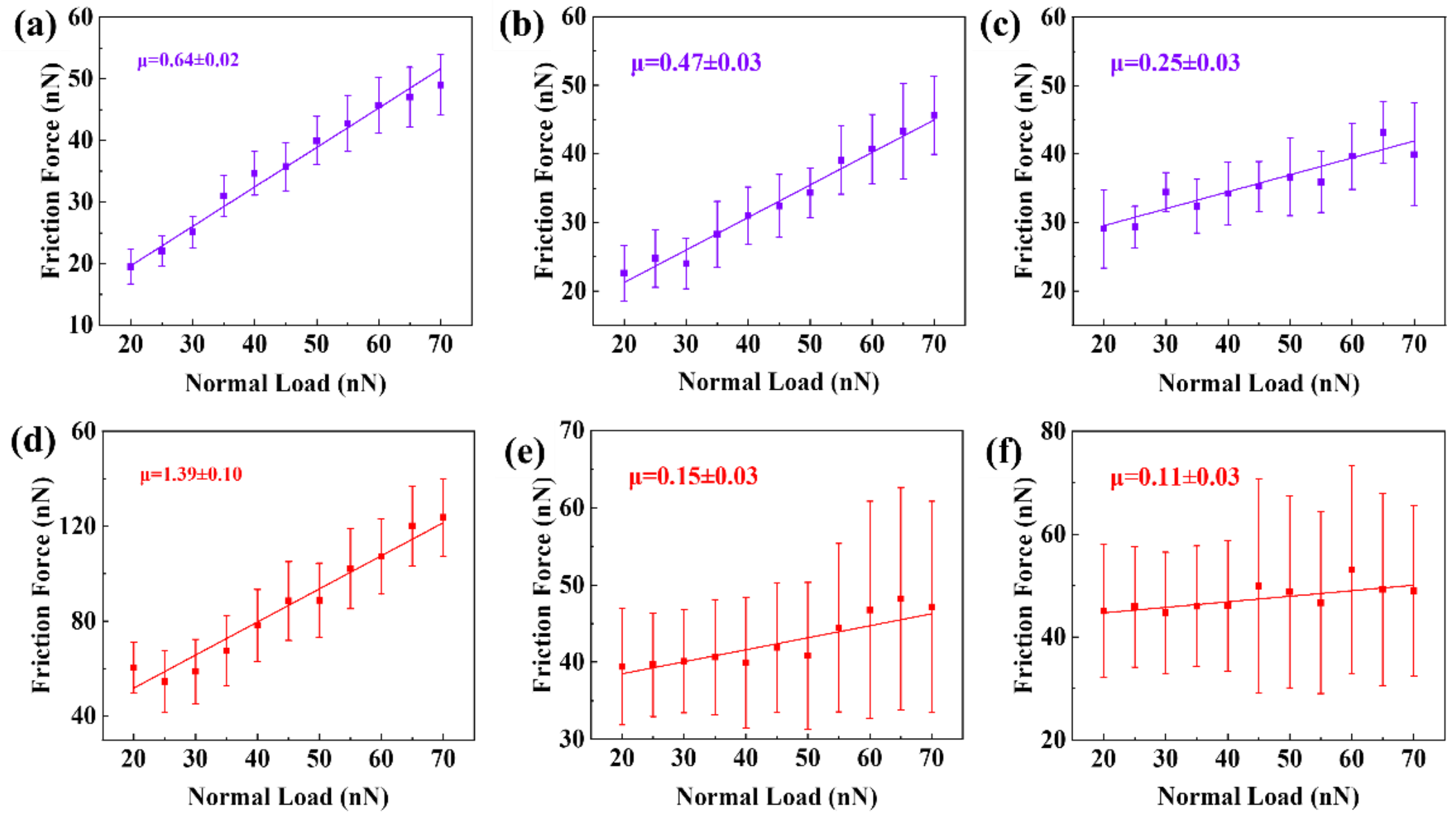

Fig. 10 Load dependence nanofriction result of $WSe_{2-x}$: (a) coating, (b) 1 N wear track, and (c) 10 N wear track; nanofriction result of $MoS_2$: (d) coating, (e) 1 N wear track, and (f) 10 N wear track.

### 3.4 The effect of variable load during the sliding

As shown above, $WSe_{2-x}$ coating shows a load dependence of the coefficient of friction typical of TMD materials, which decreases with normal load. However, there is still an open question about the mechanisms. In general, lower friction at higher loads is attributed to a better-developed (thicker, more aligned) TMD tribolayer. Consequently, if we produce the tribolayer at a high load, the friction should then stay the same when the tribological test continues at a lower load. To test this hypothesis, we performed a sliding test of $WSe_{2-x}$ coating with periodic change of load from 1 to 10 N every 200 laps, see Fig. 11. If we excluding the first 200 laps at 1 N, where elevated friction is clearly attributed to running-in, the average CoF for 10 N started at 0.056 and slowly decreased to 0.051 for the last 200 laps. For 1N, the CoF decreased slightly from 0.068 to 0.063 as well. Note that the change in coefficient of friction is abrupt when the load is adjusted in both directions, strongly suggesting that it is not related to structural/morphological changes of the tribolayer.

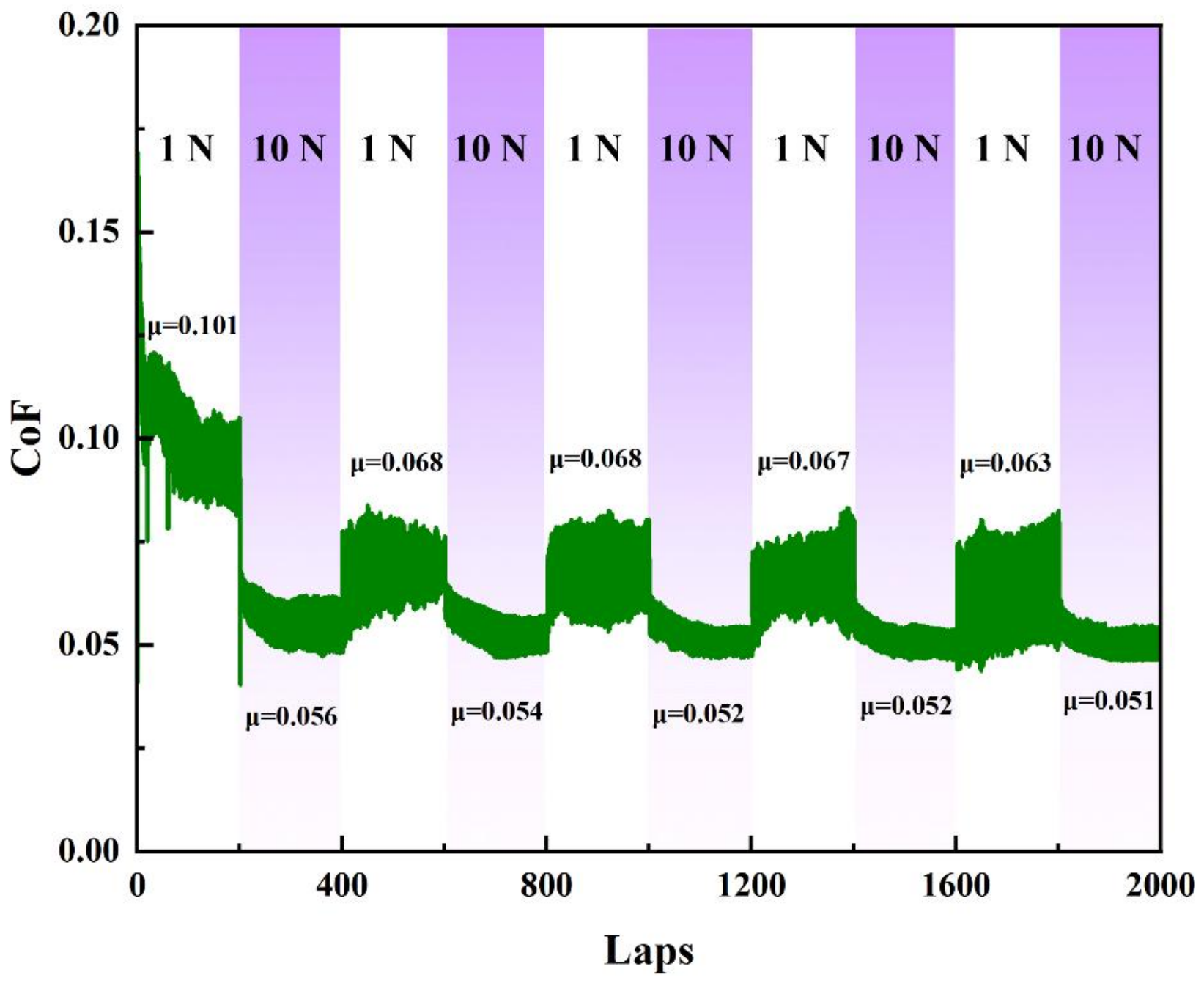

Fig. 11 Coefficient of friction of $WSe_{2-x}$ coating with periodic change of normal load between 1 N and 10 N

### 3.5 The effect of humidity

Air humidity is a crucial factor influencing the lubrication properties of TMDs by increasing interlayer shear strength [45]. High humidity leads to a high coefficient of friction, as documented for $MoS_2$ [54] and $WS_2$ [55]; however, such an effect is much weaker for diselenides [10]. We performed sliding tests in vacuum (under $7\times10^{-3}$ Pa) at 1 N and 10 N normal load (Fig. S10); the $WSe_{2-x}$ coating showed similar frictional response both in absolute value and load dependence (in vacuum, the coefficient of friction was 0.085 and 0.058 for 1 and 10 N, respectively).

## 4 Discussion

### 4.1 Friction and wear mechanisms of $WSe_{2-x}$ coating

We combine our experiments with literature to elucidate possible dominant tribological mechanisms. The nanofriction maps produced by AFM clearly show that the coefficient of friction in the wear track is much less uniform when compared to the as-deposited coating, indicating a combination of morphological and structural changes at the surface. The coefficient of friction in the wear tracks is lower in both TMDs; for $WSe_{2-x}$, the value dropped from 0.64 (free surface) to 0.47 (1N wear track) and to 0.25 (10N wear track). Although the sliding induced recrystallization and formation of well-aligned (with basal planes parallel to the surface) crystalline tribolayers on the wear track is the most likely explanation, we have to assess first other possible key players, such as oxidation and the effect of roughness.

Sliding at room temperature with our sliding speed and load range significantly reduces the detrimental effect of direct surface oxidation. Indeed, the Raman analysis of the transferred material to the balls showed only crystalline TMD material and no oxides. High roughness increases nanofriction, but in our case, the coefficient of friction decreased significantly despite a rougher wear track surface. Thus, the oxidation effect is negligible, and the change in roughness cannot explain the observed frictional behavior. Consequently, the most likely explanation is the transformation of the wearing surface from a less ordered (polycrystalline) surface to a more ordered tribolayer. Typical tribolayers of TMD-based coatings are predominantly formed through tribochemical process (diffusion-driven), which produces almost pure 2D TMD sheets at the wearing surface. Such tribolayers formation is not hindered by nonstoichiometric composition [24,50]. Contamination or alloying element(s) in the coatings, such as O [48], C [56] or N [57]) (even with very high concentrations) are removed from the contact area. Therefore, the absence of oxides in transferred material to the ball wear scar indirectly indicates the formation of such a tribolayer. Direct analysis of the wear tracks by Raman spectroscopy clearly identified higher crystallinity of $MoS_2$; however, it was much less evident in the case of $WSe_{2-x}$ due to the more pronounced initial (as-deposited) crystallinity of this coating. Nevertheless, it seems that the formation of $WSe_2$ tribolayer requires a higher energy input (provided by load) when compared to that of $MoS_2$. The drop in nanoscale coefficient of friction for $MoS_2$ is significant (about one order of magnitude) and the value is almost the same for 1 and 10 N load; for $WSe_{2-}$ $_x$, we observe a much smaller decrease in coefficient of friction in 1N wear track (0.6 to 0.47) with a further significant decrease in case of 10 N wear track (0.25). Moreover, the presence of low friction areas (Fig. 8 d) is a clear indication that the well-ordered tribolayer is formed in larger quantities only during the sliding at higher contact pressures.

Raman spectroscopy provides similar evidence that $WSe_2$ tribolayer requires higher energy input than that of $MoS_2$. $E_{1g}$ peak position for $WSe_2$ shifts significantly at the highest load (10 N), while the peak for $MoS_2$ remains relatively unchanged (Fig. 6). Additionally, the intensity ratio of $I(A_{1g})$ to $I(E_{2g})$ in Raman spectra (Fig. 7) is independent of load, suggesting that the tribolayer fully forms even at the lowest load. The same ratio for $WSe_2$ changes significantly at the highest load, corroborating nanoscale friction testing.

We can conclude here that there is strong indirect evidence of tribolayer formation in the wear track of $WSe_{2-x}$ coating, although individual techniques, such as Raman spectroscopy, provide only limited information due to the polycrystalline nature of the film. It is likely that high deficiency in selenium, resulting in high hardness, limits the formation of a low-friction tribolayer at lower loads. Thus, the drop in coefficient of friction with increasing load from 0.083 at 1 N to 0.054 at 10 N (Fig. 4 and Fig. S3) closely follows behavior observed in other TMD-based coatings [58–60], where enhanced tribolayer formation leads to reduced macroscale friction. Moreover, there is strong indirect evidence of the tribolayer formation – transfer of the coating material on the ball wear scar, clearly evident for $MoS_2$ and, to a lesser extent, for $WSe_{2-x}$ coating as well (Fig. S7). Such a transferred tribolayer is typical of TMD-based coatings [61].

Cyclic change of load resulted in reproducible oscillation of the coefficient of friction. A rapid drop (when the load increases) or rise (when the load is lowered) in friction suggests that a structural change of the wear track surface (in other words, a change of the tribolayer) is an unlikely factor explaining the sudden jump in friction. Thus, we are left with two major factors: (i) the direct effect of load on sliding, where higher compression reduces the intrinsic friction of $WSe_2$ monolayers and/or affects the interaction with water molecules, and (ii) the indirect effect due to increased energy input resulting in higher contact temperature. For the first hypothesis, there is no evidence in the literature that higher load causes a drop in friction for the genuine (with no substrate effects) sliding of 2D TMD materials. Indeed, our previous study on MoS2 friction using density functional theory suggests an increasing barrier to sliding motion (and hence friction) when the contact pressure was increased [55]; moreover, we did not observe any decrease in the coefficient of friction during sliding of homo and heterostructures with $MoS_2$ studied both experimentally (sliding of small 2D flakes) and theoretically (molecular dynamics) [56]. For the second hypothesis, we performed additional experiments to demonstrate whether higher contact temperature may reduce the friction. Fig. S11 shows the temperature dependence of the coefficient of friction. At a temperature range of 100 - 300 °C, the CoF reduced to a very low value of 0.020 - 0.025 when a load of 5N was used. Note that in this case, a silicon nitride ball was used as a counterpart to avoid rapid 100Cr6 oxidation and softening. High temperature can contribute to low friction by the reduction of sliding resistance (e.g., adhesion caused by Van der Waals forces), or by drying of the atmosphere (humidity close to the heated coating surface is negligible). However, our $WSe_{2-x}$ coating's frictional response to humidity is very low, as shown in section 3.5. Therefore, we can conclude here that, once the surface is polished and tribolayer is formed, the load-dependence of the coefficient of friction is primarily caused by increased contact temperature when the load is higher.

### 4.2 Tribological performance of $WSe_{2-x}$ coating in the context of solid lubricants

Despite a high deficiency in selenium, $WSe_{2-x}$ coating shows excellent sliding properties with a very low coefficient of friction, decreasing with higher applied loads. $WSe_{2-x}$ coating friction is not sensitive to humid air (see Fig. S10) and outperforms most of the other transition metal dichalcogenide-based coatings. To support such a statement, we compare our results with those reported in the literature for pure and doped $MoS_2$ coatings [24,57,62–70] , $WS_2$-based coatings [71–76], and $MoSe_2$ based coatings [77,78], as illustrated in Fig. 12. The values were obtained under similar conditions (sliding against a steel ball in humid air). A very low value of shear strength calculated from Equation 1 (Fig. 5), which is almost four times lower than that of $MoS_2$, further indicates the potential of tungsten diselenide as a solid lubricant.

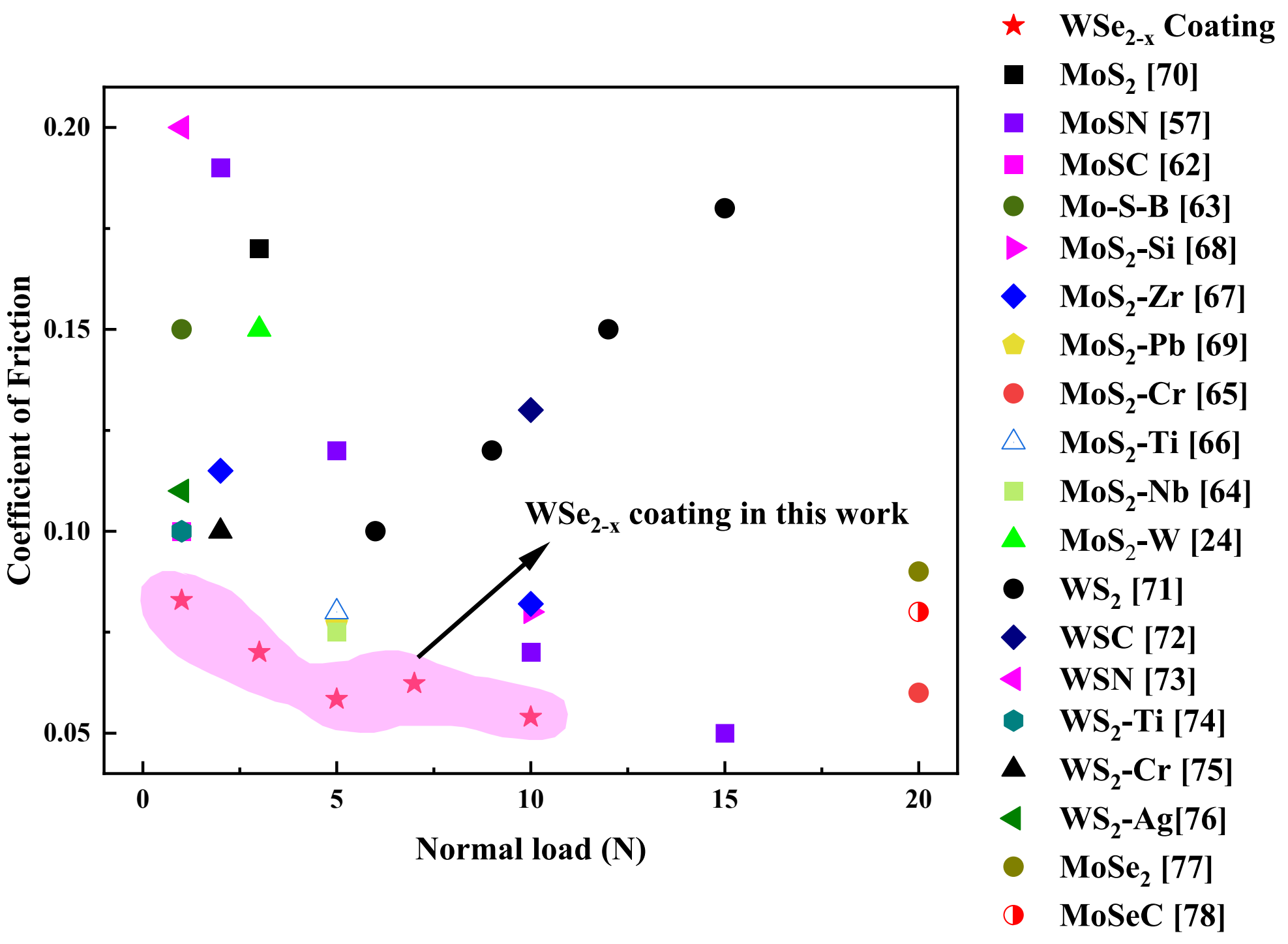

Fig. 12 The comparison of $WSe_{2-x}$ coating and other TMDs based solid lubricating coatings.

Similarly, the wear rates of $WSe_{2-x}$ coating in this study, $0.46\times10^{-6}$ to $2.20\times10^{-6}$ $mm^3$/Nm, are much lower than our reference $MoS_2$ coating or similar coatings reported elsewhere [71,79–81]. It is worth noting that, thanks to a tribolayer formation, the coefficient of friction for various coatings (e.g., $MoS_2$) is relatively comparable. However, the wear rate is much more dependent on coating hardness, density, and structure, so the direct comparison of the wear rates is only indicative.

## 5. Conclusion

This study investigated tribological properties of solid lubricant $WSe_{2-x}$ coating deposited by magnetron sputtering and compared it to well established $MoS_2$ deposited under identical conditions. The coating was tungsten-rich with a Se/W ratio of 0.92; structural analysis showed a mixture of $WSe_2$ and metallic tungsten phases, which contributed to high hardness and Young's modulus. Both $WSe_{2-x}$ and $MoS_2$ showed similar non-Amontonian tribological behavior – the coefficient of friction decreased with increasing load thanks to the formation of a well-oriented tribolayer identified by surface analyses and nanotribological tests. However, $WSe_{2-x}$ required higher energy input to form such tribolayer, likely due to a low Se/W ratio. The wear of the film was significantly lower than that of $MoS_2$, demonstrating the promising tribological potential of tungsten diselenide. Furthermore, the friction performance of the $WSe_{2-x}$ coating was similar under ambient and vacuum conditions, indicating a reduced sensitivity to oxidation and humidity.

**Acknowledgments**

This work was supported by GACR through grant 23-07785S and co-funded by the European Union under the project Robotics and advanced industrial production (CZ.02.01.01/00/22_008/0004590).

# Load-Dependent Sliding behavior of $WSe_{2-x}$ solid lubricant coating

Yue Wang[a,*], Himanshu Rai[a], Tomas Polcar[a,b,*]

a) Advanced Materials Group, Faculty of Electrical Engineering, Czech Technical University in Prague, Technicka 4, Prague 6, 16000, Czech Republic
b) School of Engineering, University of Southampton, Highfield, SO17 1BJ, Southampton, UK

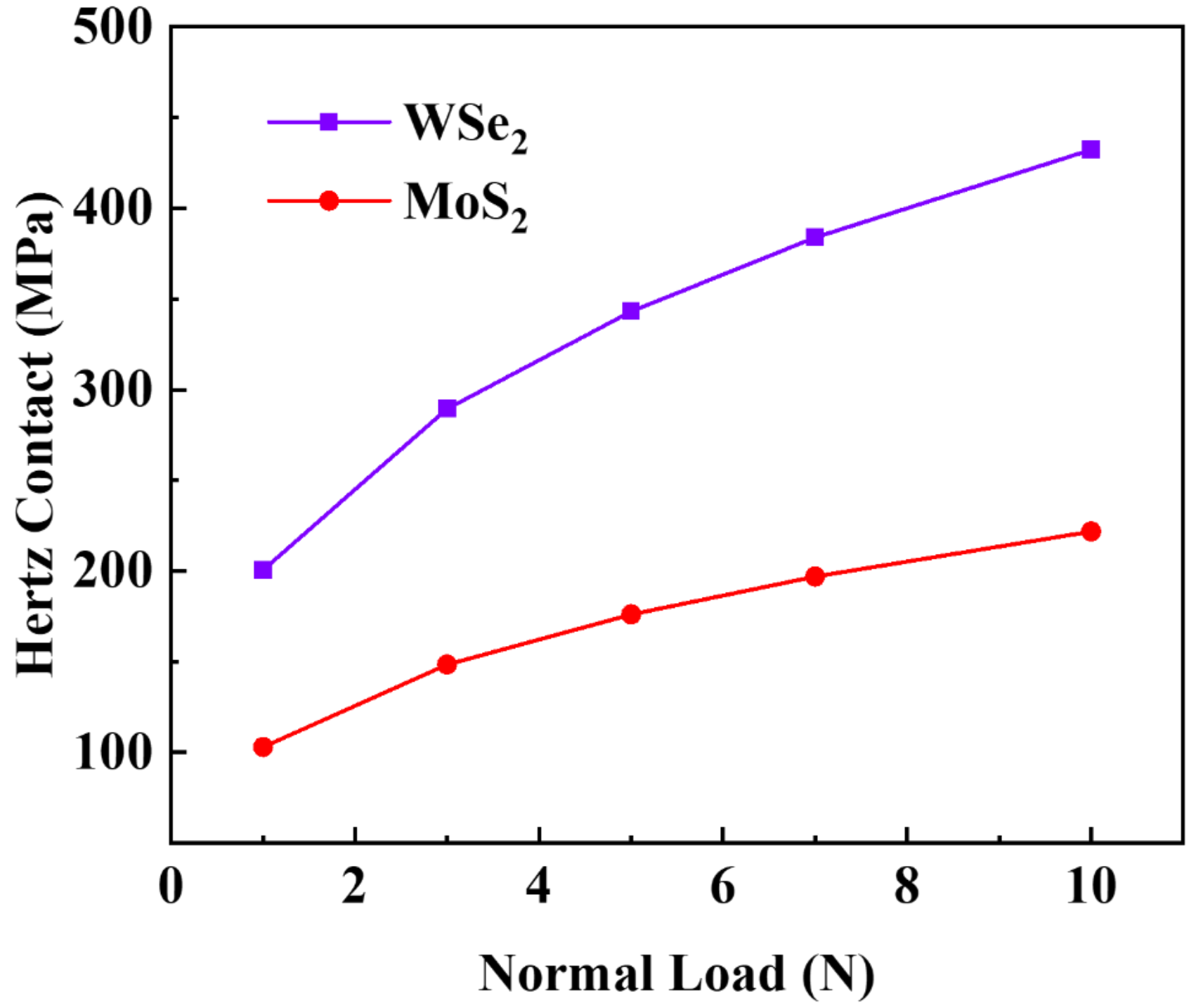

Fig. S1 Hertz contact stress with different normal load in macroscale tribological tests

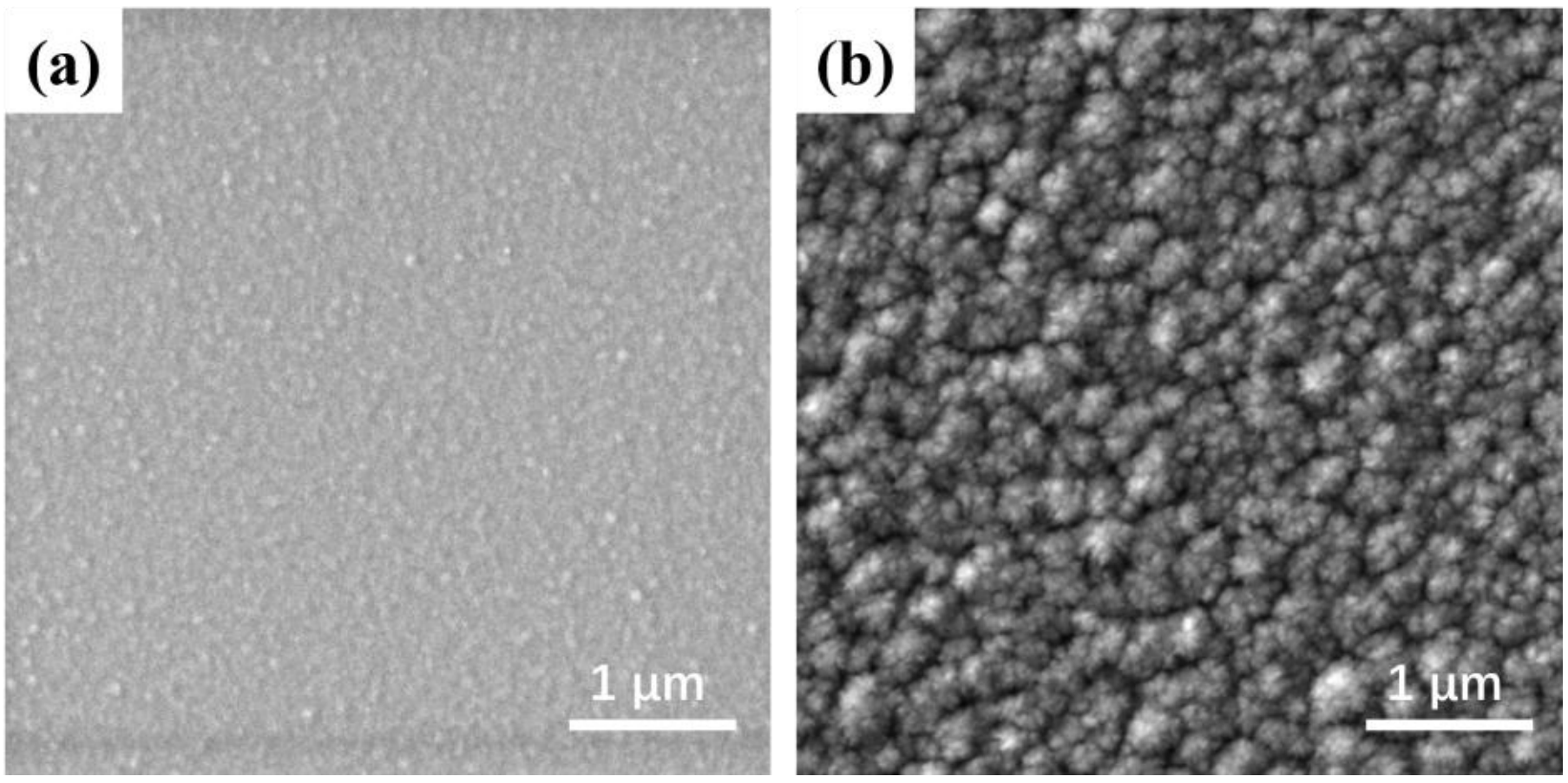

Fig. S2 Morphology of as-deposition (a) $WSe_2$ coating and (b) $MoS_2$ coating

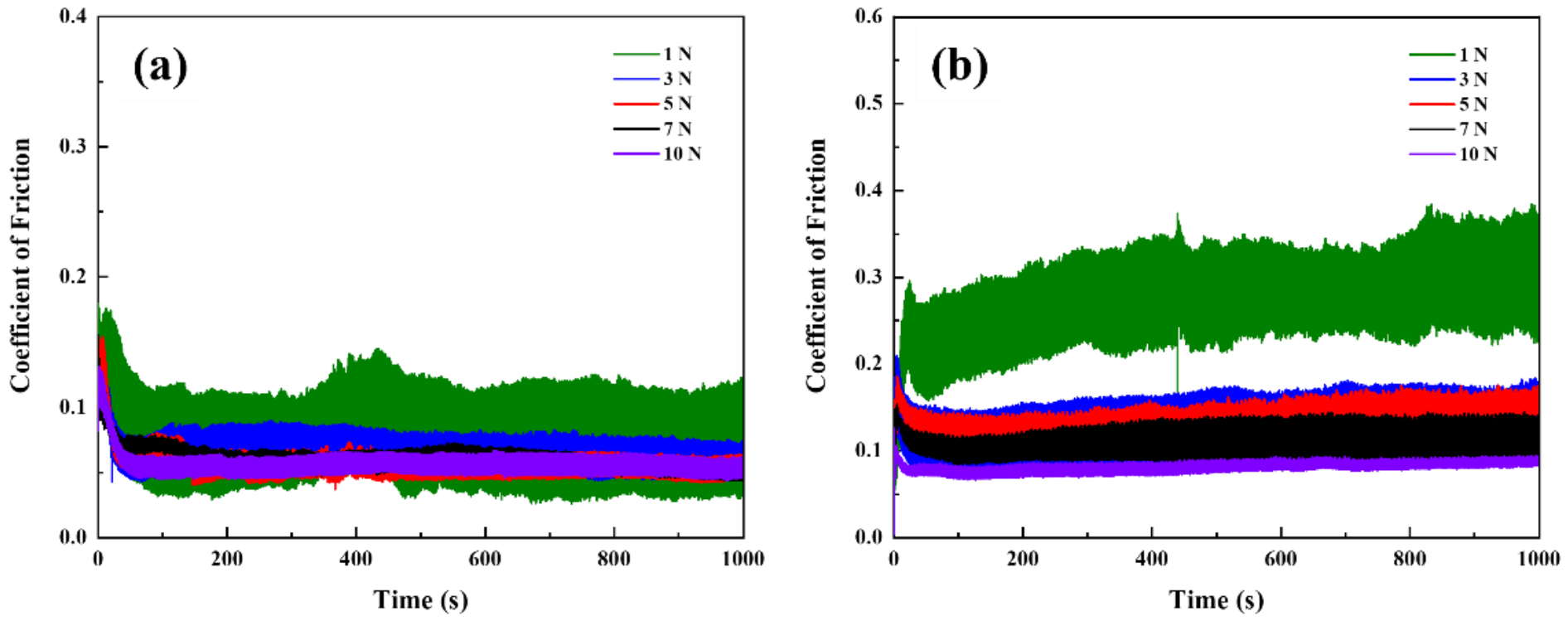

Fig. S3 (a) Friction curve of $WSe_2$ coating, (b) Friction Curve of $MoS_2$ coating

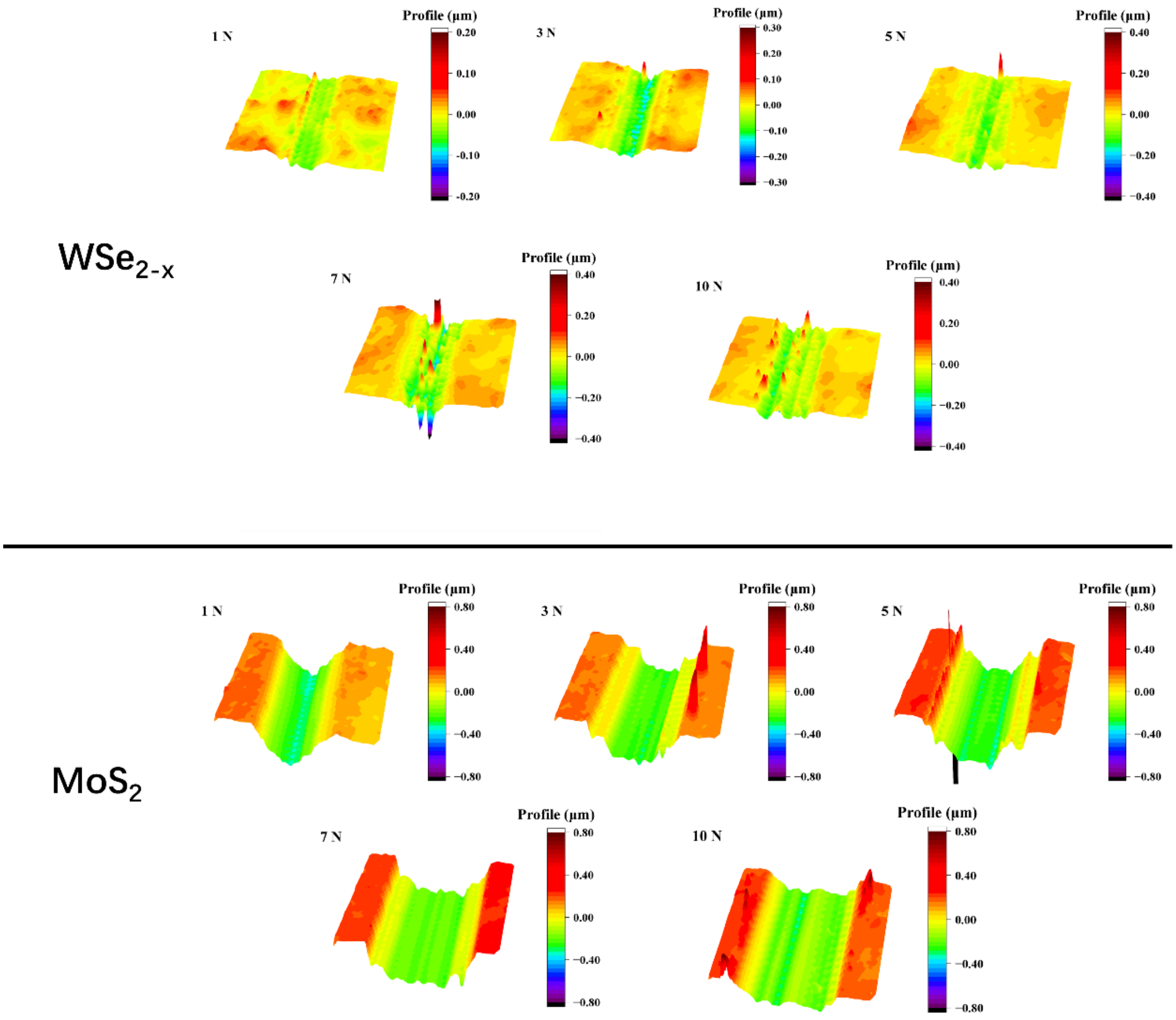

Fig. S4 Wear track morphologies on $WSe_{2-x}$ and $MoS_2$ coatings

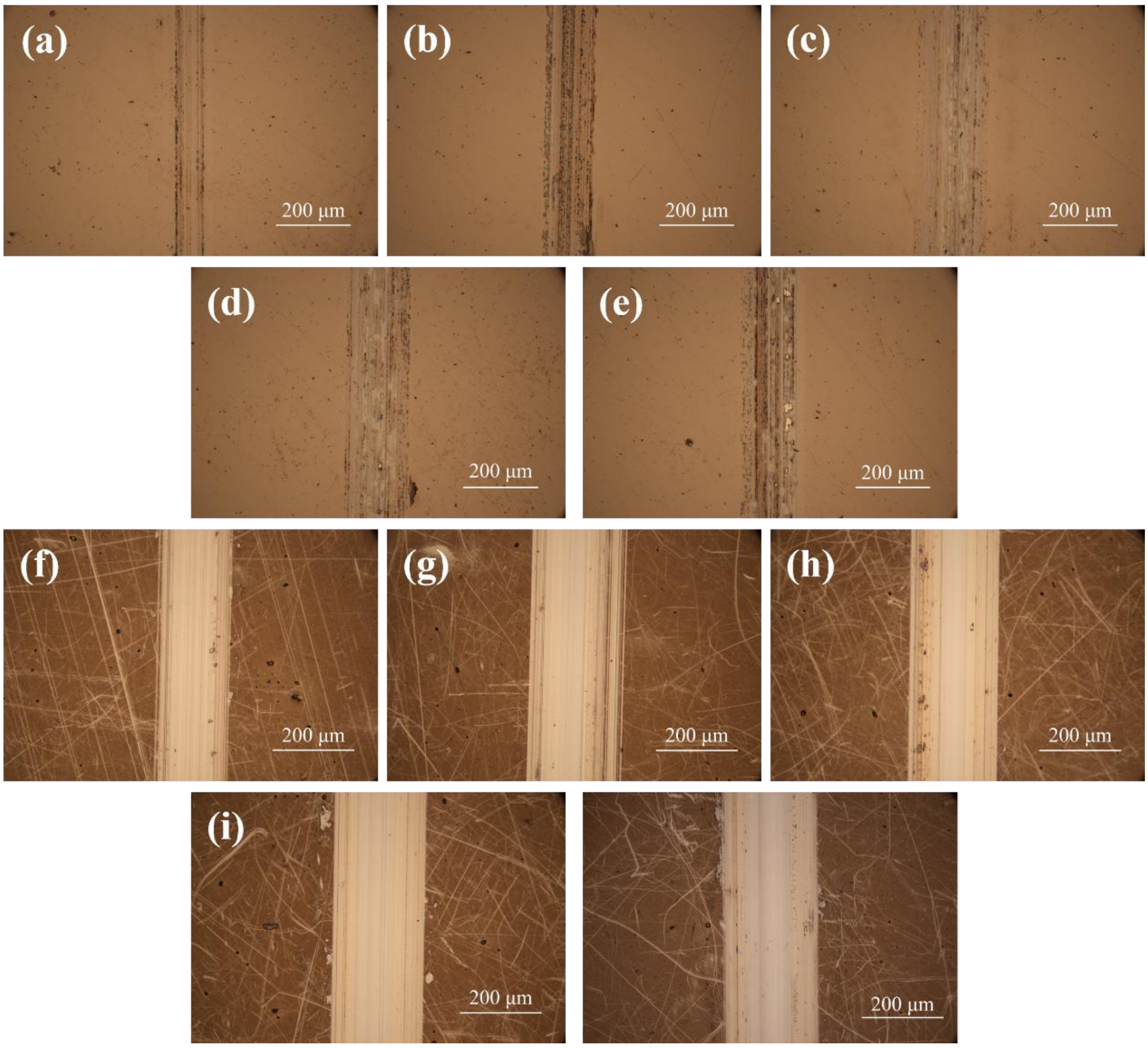

Fig. S5 Morphologies of wear tracks on $WSe_2$ resulted by (a) 1 N, (b) 3 N, (c) 5 N, (d) 7 N and (e) 10 N normal load; and the morphologies of wear tracks on $MoS_2$ resulted by (f) 1 N, (g) 3 N, (h) 5 N, (i) 7 N and (j) 10 N normal load

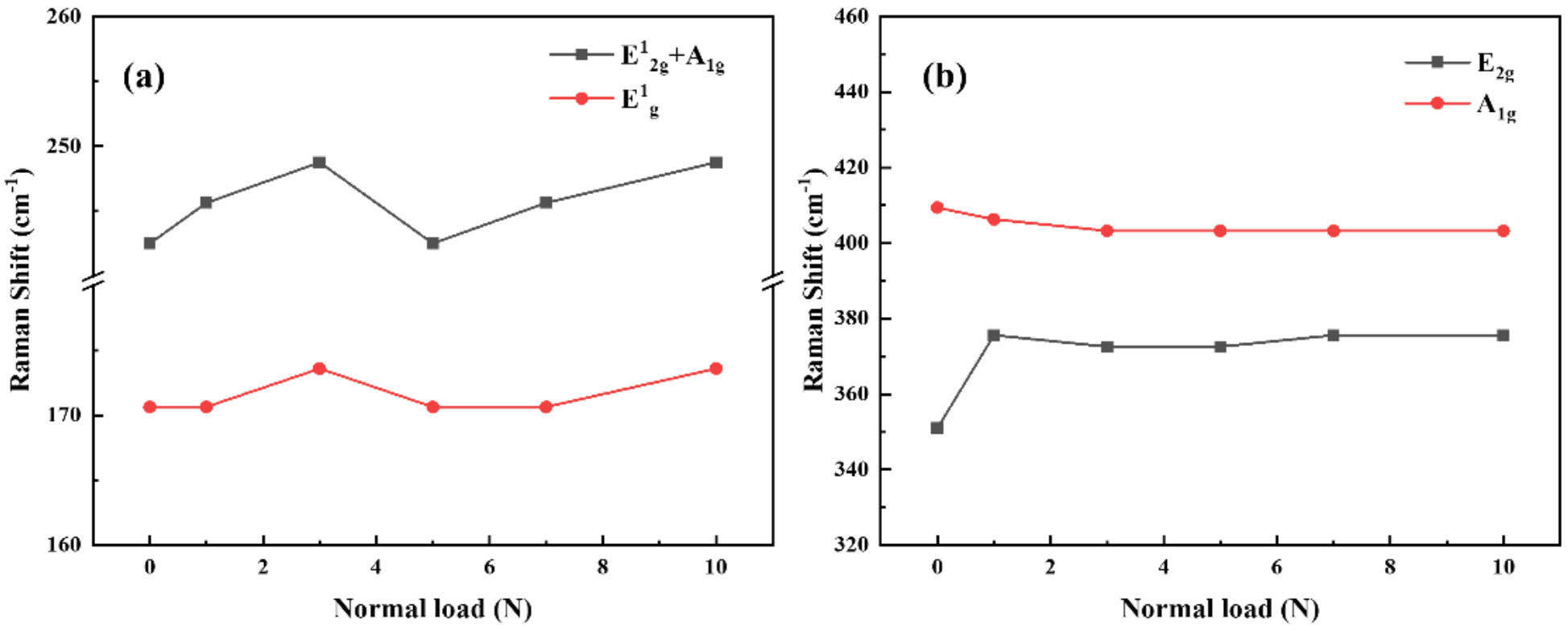

Fig. S6 Peaks position on wear tracks for (a) $WSe_2$ coating and (b) $MoS_2$ coating

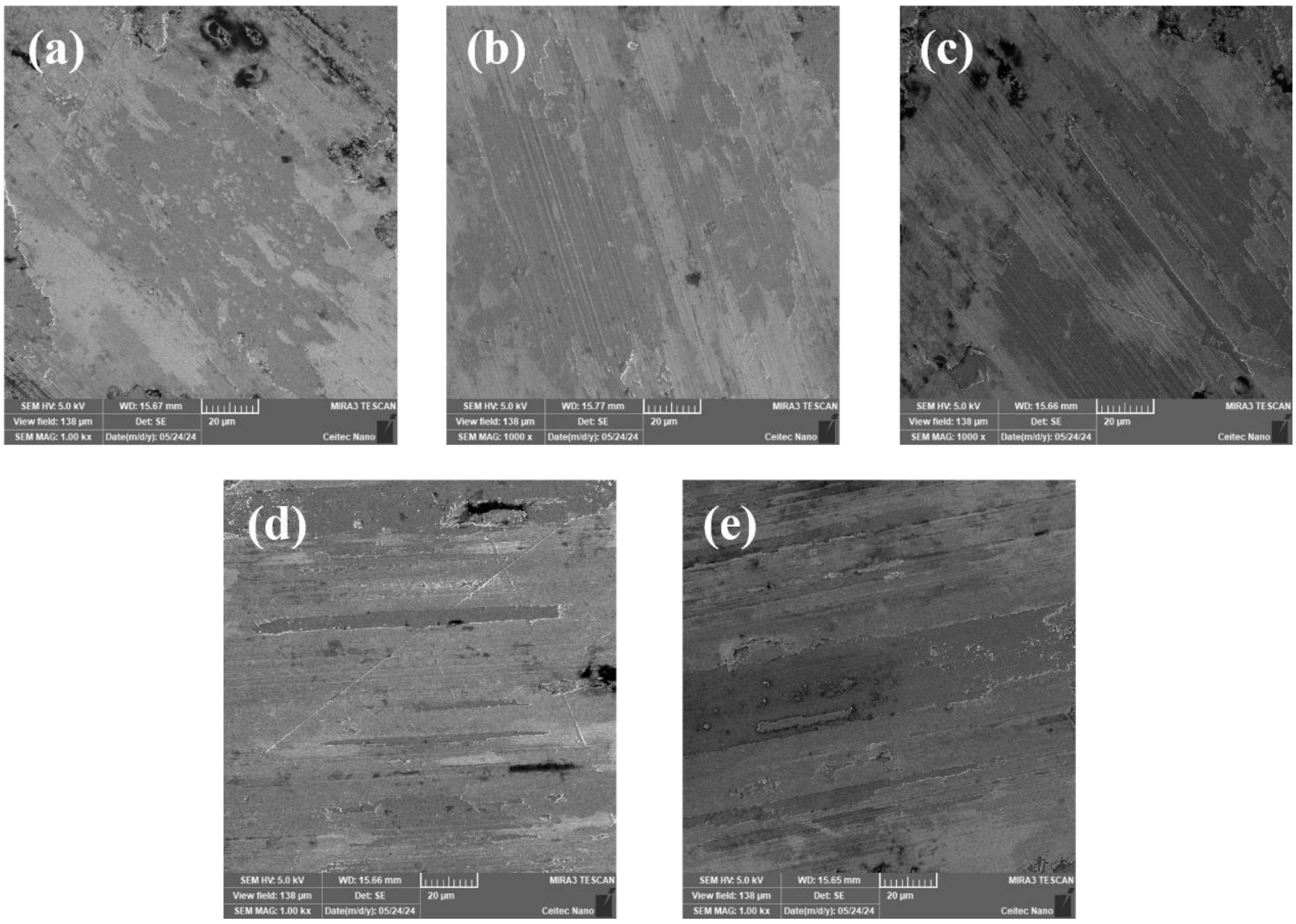

Fig. S7 Ball wear scars covered by transferred coating $WSe_{2-x}$ material: (a) 1N, (b) 3 N, (c) 5 N, (d) 7 N, and (f) 10 N

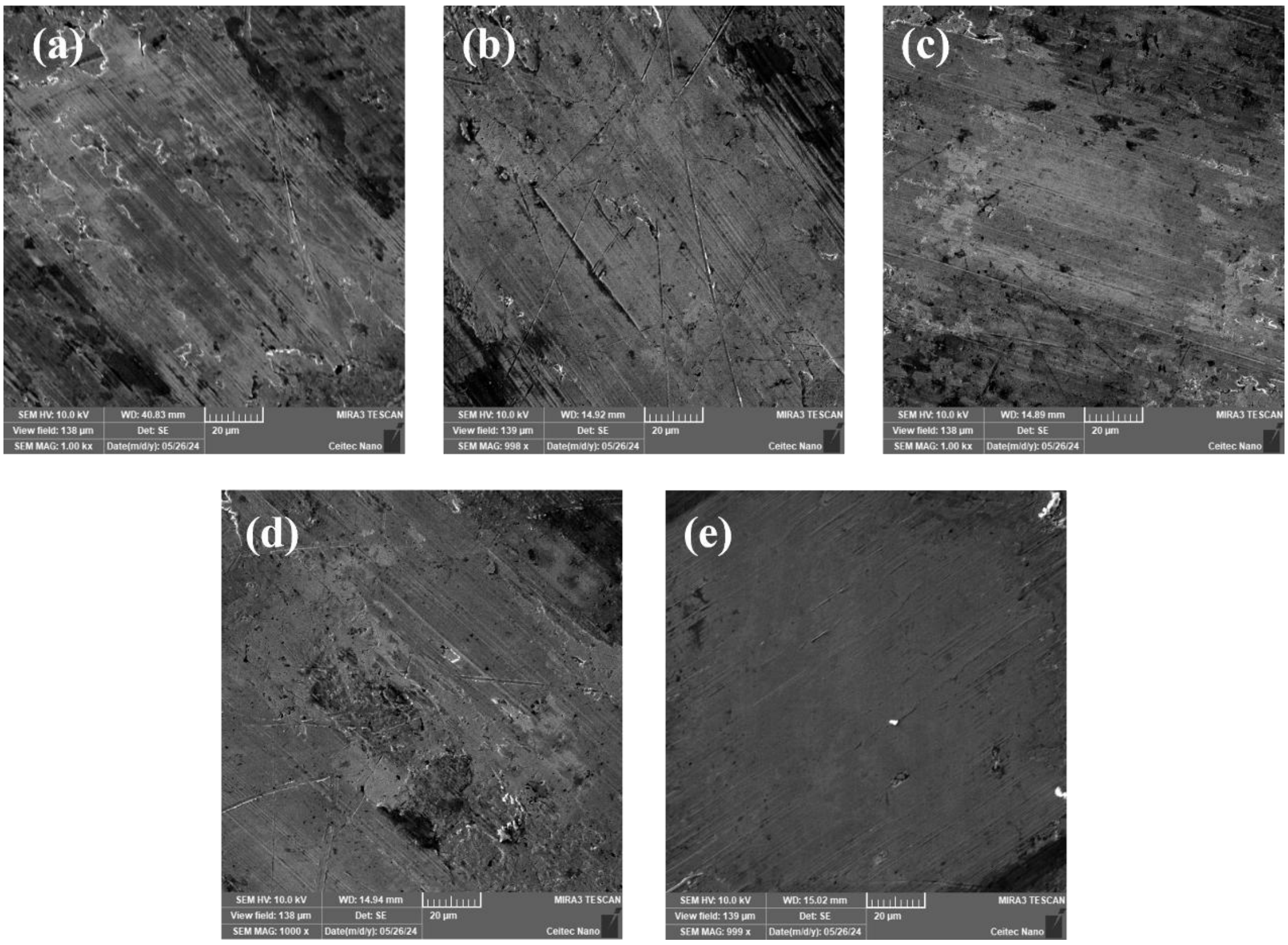

Fig. S8 Ball wear scars covered by transferred coating $MoS_2$ material: (a) 1N, (b) 3 N, (c) 5 N, (d) 7 N, and (f) 10 N

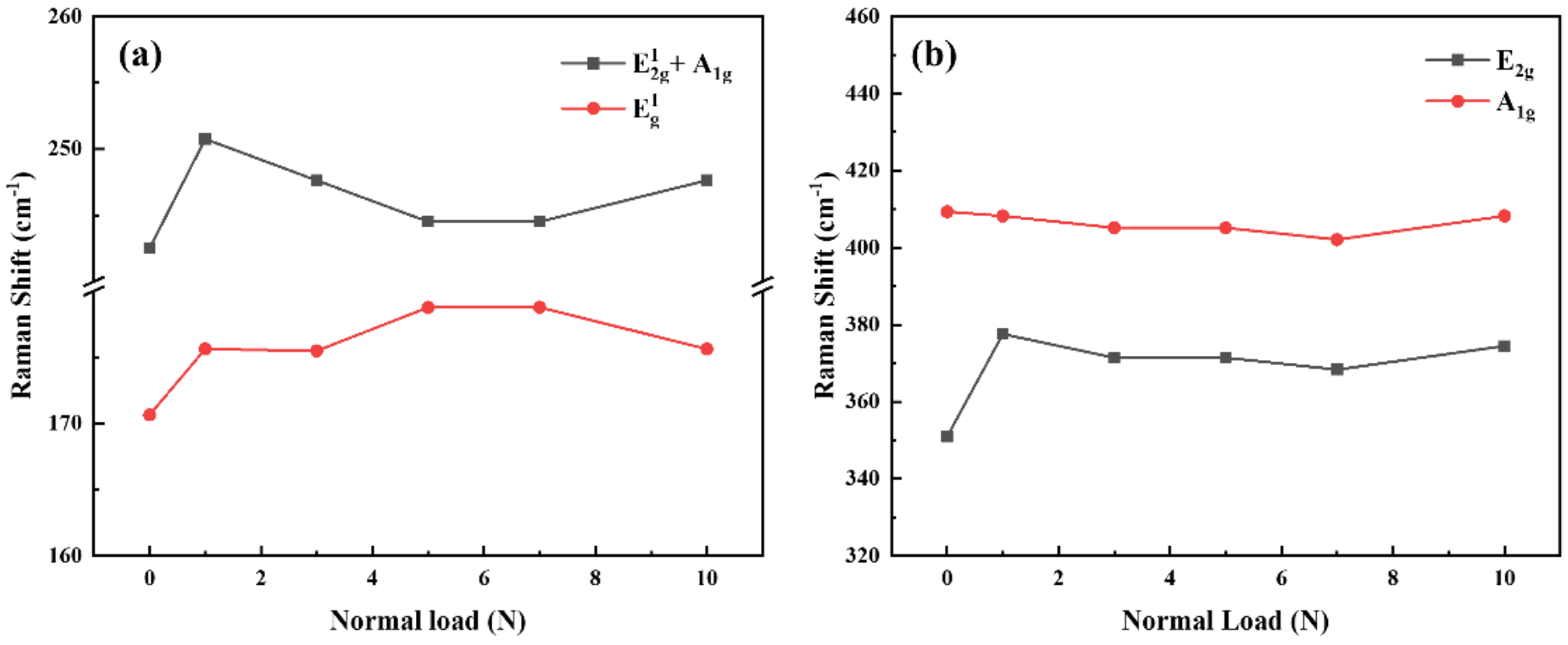

Fig. S9 Peaks position of (a) $WSe_2$ and (b) $MoS_2$ on the ball scars

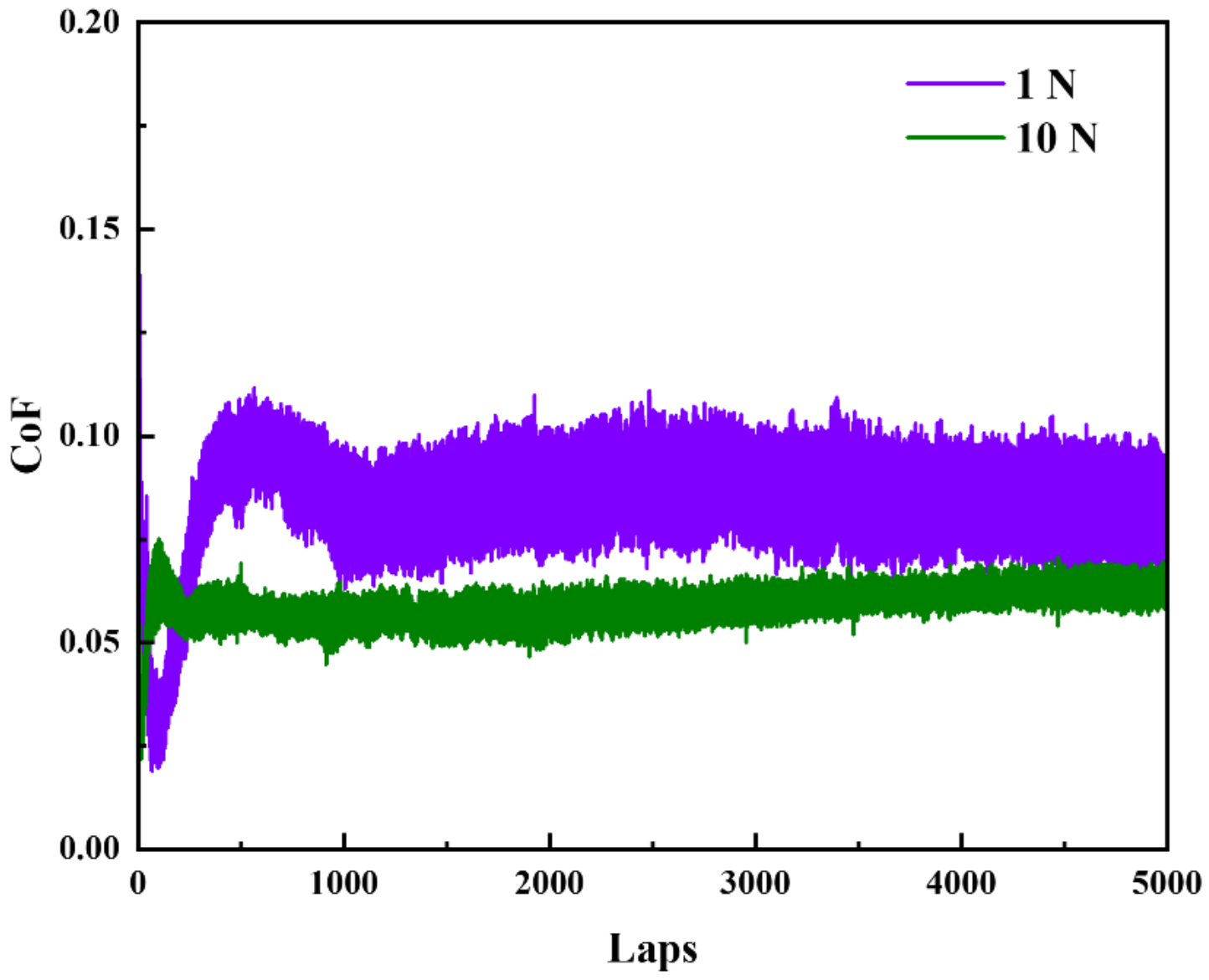

Fig. S10 Vacuum sliding test on $WSe_{2-x}$ coating under $7\times10^{-3}$ Pa.

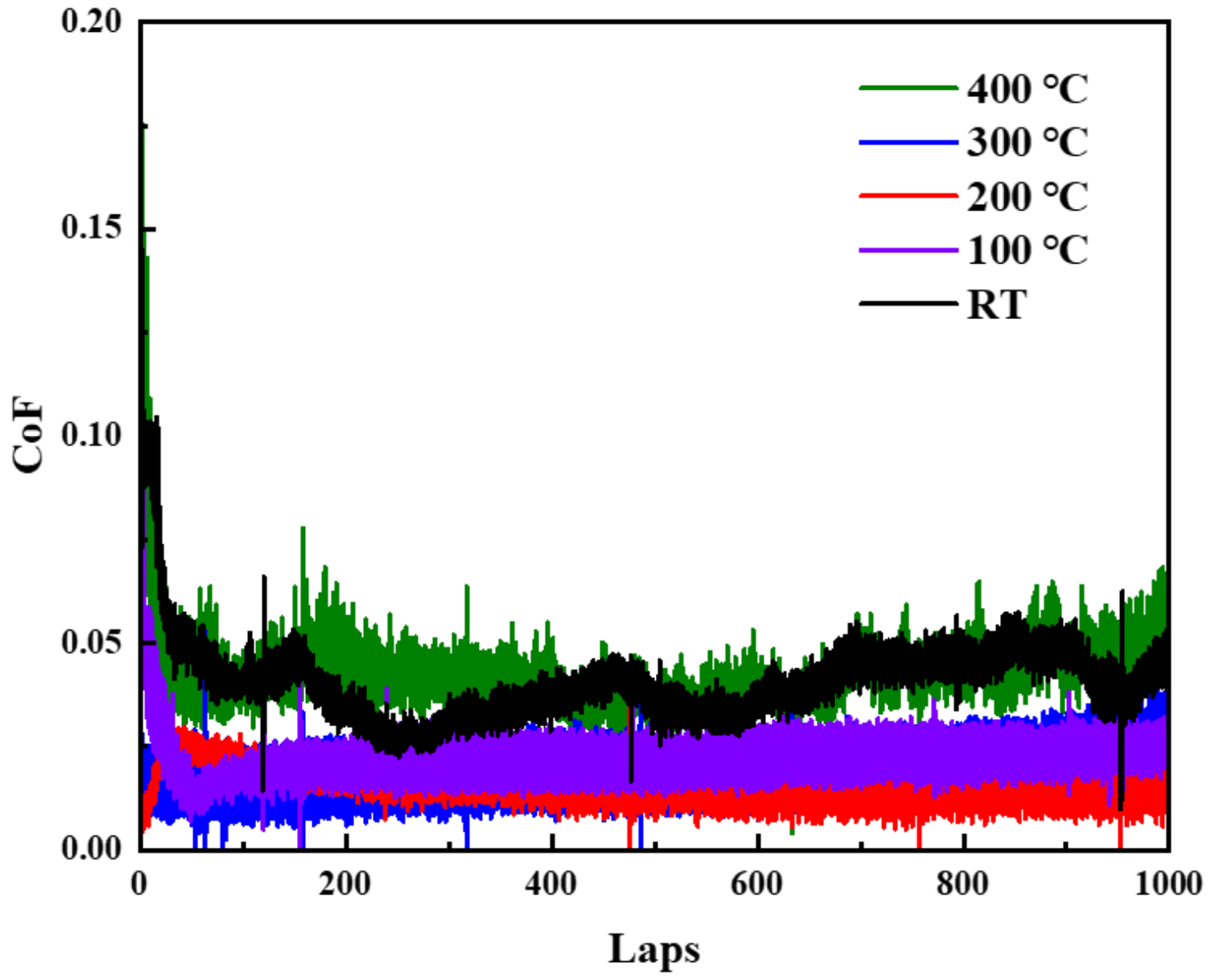

Fig. S11 Temperature dependence sliding behavior of $WSe_{2-x}$ coating from room temperature to 400 °C.